\documentclass{jssc}
\newenvironment{myproof}{{\noindent \textbf{Proof:}}}{\hfill $\square$ \par}
\def\textsubscript#1%
{$_{\text{#1}}$}

\def\cdd{\mbox{\boldmath$\cdot$}~}
\input{amssym.def}
\include{graphix}

\makeatletter
\def\@oddfoot{\hfill}
\newcount\shumeicount
\def\setshumei#1#2#3{%
  \shumeicount=\count0
  \def\@oddhead{%
    \raise-5pt\hbox to0pt{\vrule width\hsize height 0pt depth 0.4pt\hss}\relax
    \ifnum \shumeicount=\count0
      \raise-7pt\hbox to0pt{\vrule width\hsize height 0pt depth 0.4pt\hss}\relax
      #1
    \else
      \ifodd\count0
        #2
      \else
        #3
       \fi
     \fi
  }%
}
\makeatother
\makeatletter
\def\@oddfoot{\hfill}
\newcount\shujiaocount
\def\setshujiao{%
  \shujiaocount=\count0
  \def\@oddfoot{%
      \ifodd\count0
      \else
      \fi
  }%
}
\makeatother
\def\title#1#2#3#4{{
  \vspace*{0.3cm}
  \begin{flushleft} \Large\bf #1\end{flushleft}
  \vspace*{-0.2cm}
      \begin{flushleft}
      \bf #2
      \end{flushleft}
      \footnotetext{\hspace{-6mm} #3\\ #4}}}
\def\dshm#1#2#3#4
{\setshumei{ (#1) #2:
{\thepage--\pageref{LastPage}}\hfill}
            {\hfill {\small #3}\hfill\hbox to0pt{\hss\thepage}}
            {\hbox to0pt{\thepage\hss}\hfill {\small #4}\hfill
            }
            \setshujiao}
\def\drd#1#2
{{\vskip 1cm\small \begin{flushleft}
 #1 \\
 #2\\
\copyright The Editorial Office of JSSC \&  Springer-Verlag GmbH
Germany 2021
\end{flushleft}}}
\def\hat{\widehat}

\def\epsilon{\varepsilon}

\begin{document}

%*************************************************************************************************************
% \biaoti{THE CAPITALIZED TITLE OF YOUR ARTICLE$^*$}{The list of authors' names with the LAST NAME capitalized
% and the authors' names should be separated by "\cdd"}{the first author's name \\ the first author's affiliation
% and Email address\\ the second author's name\\ the second author's affiliation. More can be listed like this.}
% {$^*$ The titles and numbers of the foundations that support this article.}
%*************************************************************************************************************
\title{Equalizer zero-determinant strategy in  discounted repeated Stackelberg asymmetric game $^*$}%%%   Main Title of your paper  %%%
{\uppercase{Cheng} Zhaoyang \cdd \uppercase{Chen}
Guanpu \cdd \uppercase{Hong} Yiguang}%%% The names of the authors  %%%
{\uppercase{Cheng} Zhaoyang \\
Key Laboratory of Systems and Control, Academy of Mathematics and Systems Science, Beijing, 100190, China, and School of Mathematical Sciences, University of Chinese Academy of Sciences, Beijing, 100049, China. Email: chengzhaoyang@amss.ac.cn\\   % Academy of Mathematics and Systems Science, Chinese Academy of Sciences, Beijing $100190$, China
\uppercase{Chen} Guanpu   \\
School of Electrical Engineering and Computer Science, KTH Royal Institute of
Technology, Stockholm, 100 44, Sweden.  Email: guanpu@kth.se \\
\uppercase{Hong} Yiguang (Corresponding author)\\
Department of Control Science and Engineering, Tongji University, Shanghai, 201804, China, and Shanghai Research Institute for Intelligent Autonomous
Systems, Tongji University, Shanghai, 210201, China. Email: yghong@iss.ac.cn\\   % Academy of Mathematics and Systems Science, Chinese Academy of Sciences, Beijing $100190$, China
   } %%% The address of the authors  %%%
{$^*$This work was supported by the National Key Research and Development Program of China under No. 2022YFA1004700, the National Natural Science Foundation of China under No. 62173250, and Shanghai Municipal Science and Technology Major Project under No. 2021SHZDZX0100.
}

%*************************************************************************************************************
%The submission date of your article. For example: \drd{Received: June 8, 2006}
%*************************************************************************************************************
\drd{DOI: }{Received: x x 20xx}{ / Revised: x x 20xx}

%*************************************************************************************************************
% The page header of the article.
% \dshm{Year}{Volume}{The capitalized RUNNING HEAD of your article with less than 48 letters}{The capitalized
% AUTHORS list with $\cdot$ separating different names or one can type "The name of the first author et al."
% if there are more than 4 authors.}
%*************************************************************************************************************

\dshm{20XX}{XX}{A TEMPLATE FOR JOURNAL}{}

%*************************************************************************************************************
% \dab{The abstract}{Keywords}
%*************************************************************************************************************
%-------------------------------------------------------------------------
\Abstract{This paper focuses on the performance of equalizer zero-determinant (ZD) strategies in discounted repeated Stackerberg asymmetric games. In the leader-follower adversarial scenario, the strong Stackelberg equilibrium (SSE) deriving from the opponents' best response (BR), is technically the optimal strategy for the leader. However, computing an SSE strategy may be difficult since it needs to solve a mixed-integer program and has exponential complexity in the number of states. To this end, we propose to adopt an equalizer ZD strategy, which can unilaterally restrict the opponent's expected utility. We first study the existence of an equalizer ZD strategy with one-to-one situations, and analyze an upper bound of its performance with the baseline SSE strategy. Then we turn to multi-player models, where there exists one player adopting an equalizer ZD strategy. We give bounds of the sum of opponents' utilities, and compare it with the SSE strategy. Finally, we give simulations on unmanned aerial vehicles (UAVs) and the moving target defense (MTD)  to verify the effectiveness of our approach.
}      % the abstract

        % the keywords

%\MRSubClass{05B05, 05B25, 20B25}      % MR(2000) Subject Classification

%\baselineskip 15pt

\section{Introduction}

\iffalse
This letter focuses on zero-determinant (ZD) control for obtaining an equalizer strategy in discounted repeated asymmetric games. Consider the leader-follower adversarial scenarios in cyber-physical systems (CPS). 
 Although the strong Stackelberg equilibrium (SSE), derived from the feedback control of opponents' best response (BR) strategy, is technically the optimal strategy for the defender,  computing an SSE strategy may be difficult due to historical information storage and computation complexity. To this end, we turn to an equalizer ZD strategy, which can be regarded as open-loop control to unilaterally restrict the opponent's expected utility. Concretely, we study the existence of an equalizer ZD strategy in the discounted repeated asymmetric game scenarios, and analyze upper bounds of its performance with the baseline SSE strategy. Then we extend the approach into multi-player game-theoretical models. Also, we give a simulation with unmanned aerial vehicles (UAVs) to verify the effectiveness of our approach.

\fi

%In recent years, more and more work has considered interactions among agents in smart grids, sensor networks, or cyber-physical systems (CPS)\cite{franze2020resilient,ding2022fully,lu2022networked,ye2022distributed,nguyen2019deception}. 

In recent years, agent interaction has been widely considered in smart grids, sensor networks, or cyber-physical systems (CPS), while many of them are modeled by game theory
\cite{liu2023optimal,chen2021distributed,umsonst2020nash,xu2023algorithm}.
Actually, agents often face the situation with dynamic and persistent interactions. For example, smart power grids often confront long-term and persistent attacks, and sensor networks also have repeated periodic detections \cite{miao2013stochastic,zhang2018dynamically}. The repeated game is a typical theoretical model to analyze the interaction in these situations \cite{mishra2020model,feng2017signaling}.

%These problems are also often modeled as repeated games over a long period. 

Actually, one player may have advantages in the strategy implementation sequence, such as the defender in moving target defense (MTD) problem and the drone leader in unmanned aerial vehicles (UAVs) \cite{li2020spatial,tahir2019swarms}. The typical case is always denoted as the repeated Stackelberg asymmetric game, which is one of the important categories to characterize players' behaviors \cite{vorobeychik2012computing,korzhyk2011stackelberg,cheng2023zero}. Specifically, consider a Stackelberg game model between two players. The follower tends to choose the best response (BR) strategy after observing the strategy of the leader, while the leader picks the strong Stackelberg equilibrium (SSE) strategy based on the opponent's  BR strategy \cite{vasal2020stochastic,9722864,lopez2022stationary}. 
Thus, an SSE strategy can be regarded as the optimal solution in a long-period repeated Stackelberg game. 

However, the computation of the SSE in such a repeated game is complex, which is always transformed into a mixed-integer non-linear program with a non-convex optimization objective \cite{lopez2022stationary,vorobeychik2012computing}. The non-convexity makes the computation much difficult, and even mixed-integer polynomial programs have exponential complexity in the size of states \cite{basu2023complexity,basu2022complexity}.  %Besides,  the learning processes from the BR strategy yield data storage for historical information, as well as plenty of time cost. 
Thus, although the SSE strategy is optimal for the leader, we hope to adopt another efficient strategy, in order to avoid the cost in computing.

\iffalse
repeated security games attract more and more attention in many fields such
as the cyber-physical system (CPS), the unmanned aerial vehicle (UAV), and  the moving target defense (MTD) \cite{8485952,feng2017signaling,kovtun2022reliability,wu2022game}. The repeated Stackelberg asymmetric security game is one of the important categories to characterize players' behaviors when the defender faces persistent threats from the attacker.As a fundamental model discussed in the previous works \cite{vorobeychik2012computing,korzhyk2011stackelberg}, the attacker tends to choose the best response (BR) strategy after observing the defender' strategy, while the defender aims to maximize its utility considering the attacker. Actually, the defender, as a leader, has an advantage in guiding the attacker’s decision, and the defender picks the optimal strategy based on predicting the attacker's BR strategy. The corresponding equilibrium is defined as the strong Stackelberg equilibrium (SSE) \cite{vorobeychik2012computing,li2019cooperation,9722864}.
\fi

%A strong Stackelberg equilibrium (SSE) strategy is important between player $1$ and B, and can be regarded as a feedback control in the actual repeated game.

%, has an advantage in guiding the attacker’s decision, and the defender picks the SSE strategy based on predicting the attacker's BR strategy. 

Markedly, the zero-determinant (ZD) strategy has become popular in repeated games \cite{press2012iterated,govaert2020zero,tan2021payoff,cheng2023zero}, which is derived from Iterated Prisoner’s Dilemma (IPD). Briefly, ZD strategies can unilaterally enforce the two players’ expected utilities subjected to a linear
relation. Besides, the equalizer ZD strategy is one of the typical classes of  ZD strategies to unilaterally set the opponent’s
utility. 
It has been widely studied to promote cooperation or unilaterally extortion in many research fields, like public goods games (PGG), human-computer interaction (HCI), and evolutionary
games \cite{wang2016extortion,hilbe2013evolution}, where most players has symmetric payoffs.

Further, the asymmetric situation has become increasingly attractive since players may have different preferences or abilities in achieving their purposes \cite{hirai2013existence}. For example, in security games, defenders often need to defend against the intrusion of many attackers of different types. In fact, the situation where each player has two actions to choose from is an important multi-player game. For example, in asymmetric public goods games and asymmetric snowdrift games, with \cite{nockur2021different,reeves2017asymmetric,du2009asymmetric}, players can choose to cooperate and work together to complete the task, or refuse to make an effort and only enjoy the benefits. As an extension of repeated symmetric games \cite{govaert2020zero}, the discussion of ZD strategies and SSE strategies in such an asymmetric game is also important.

\iffalse
However, ZD strategies in repeated multi-player games are still important, although most literature focuses on two-player games. For example, researchers have studied the existence of ZD strategies in multi-player public goods games. At the same time, the characteristics of viable ZD strategies in multi-player social dilemmas, as well as strategies for maintaining cooperation in such multi-player games, have been extensively studied. Of course, in multiple repeated games, some constraints are necessary to ensure the existence of a viable ZD strategy.

\fi

Hence, we are inspired to apply ZD-based approaches to a class of repeated games, which is representative and important in CPS or UAV scenarios. We consider discounted long-term utilities in asymmetric situations, which are consistent with the fact that players are attracted by the reward in recent stages and may have individual preferences. %Besides, the game in actual interactions is not symmetric. For example, the preferences of attackers and defenders are often opposite in CPS.
On this basis, this paper studies the performance of the equalizer ZD strategy for the leader, compared with the baseline SSE strategy. The main contribution of this work is summarized as follows.

\begin{itemize}
\item In the discounted repeated Stackelberg asymmetric game with one-to-one scenarios, we reveal an existence condition of equalizer ZD strategies. Also, we analyze an upper bound of expected utilities when adopting an equalizer ZD strategy. Our results give the leader a choice set in selecting the ZD strategy to avoid computation burden in seeking the SSE strategy.

\item We further study multi-player scenarios, and similarly, when one player chooses an equalizer ZD strategy, we give a bound of the sum of opponents' utilities. Also, in the multi-player situation, we show the gap between the leader's utility with ZD strategies and that with SSE strategies. The leader could adopt the equalizer ZD strategy and maintain its utility, facing multiple opponents.

\item We verify our results in experiments by providing the leader with proper equalizer ZD strategies and making comparisons with an SSE strategy as the baseline. In the moving target defense (MTD)
and unmanned aerial vehicles (UAVs), we show utility performances and action interactions among players. Our experiments also illustrate that the equalizer ZD strategy can help the leader maintain its utility.

\end{itemize}

This paper is organized as follows. In Section 2, multi-player game models and strategies are provided. In Section 3, we consider a typical case, the one-to-one situation, and show the utility analysis between ZD strategies and SSE strategies. Also, in Section 4, we extend the  results in one-to-one scenarios to multi-player scenarios, which are more general than one-to-one
models in real stochastic adversarial scenarios. In Section 5, we show the experiments to verify our results in 
both the one-to-one situation and the multi-player situation. Finally, we conclude this article in Section 6.

\section{Preliminary}

In this section, we give the formulation of the multi-player repeated game, and show the definition of strong Stackelberg equilibrium (SSE) strategies and zero-determinant (ZD) strategies.

\subsection{Repeated leader-follower game}

Consider a repeated asymmetric game $\mathcal{G}=\{N,\mathcal{A},\mathcal{S},r,P\}$.   $N=\{1,\dots,n\}$ is the player set. Each player has two actions $\mathcal{A}_i=\{1,2\}$ to select in each stage, and $\mathcal{A}=\mathcal{A}_1\times\dots\times\mathcal{A}_n$ is the action set of all players. $\mathbf{r}=\{r_1,r_2,\dots,r_n\}$ is the reward set of players. For convenience, at each stage, denote $1$-selector as the player chooses action $1$, and $2$-selector as the player chooses action $2$.  %Actually, $r_i(a,b)=U_{ab}^i$ for $a\in\mathcal{A}$ and $b\in\mathcal{B}$.
 Each player's utility depends on its own action and the number of $1$-selectors among other players, as shown in Table \ref{tab::multi}. For example, for player $i$ facing with the situation where there are $k$ players selecting action $1$, player $i$ gets $U^i_{1,k}$ when it chooses action $1$, and gets $U^i_{2,k}$ when it chooses action $2$. We consider that the utility matrix is asymmetric for players, i.e., $U^{i_1}_{jk}\neq U^{i_2}_{jk}$, for some $i_1\neq i_2$, $j\in \{1,2\}$, and $k\in\{0,1,\dots,n-1\}$.

\begin{table}
\renewcommand\arraystretch{1.3}
\tabcolsep=0.2cm
\caption{Utility matrix of player $i$}  
\centering  

\begin{tabular}{ccccccc}
\hline $\begin{array}{c}\text { Number of $1$-selectors } \\
\text { among other players }\end{array}$ & $n-1$ & $n-2$ & $\cdots$  & 1 & 0 \\
\hline $1$-selector's payoff & $U^i_{1,n-1}$ & $U^i_{1,n-2}$ & $\cdots$  & $U^i_{1,1}$ & $U^i_{1,0}$ \\
$2$-selector's payoff & $U^i_{2,n-1}$ & $U^i_{2,n-2}$ & $\cdots$ &  $U^i_{2,1}$ & $U^i_{2,0}$  \\
\hline
\end{tabular} 
       \label{tab::multi}  
\end{table}

\begin{example}\label{ex::3.1}

The asymmetric utility matrix in Table \ref {tab::multi} is an extension of the matrix in some typical symmetric games. Actually, consider the case that $U^{i_1}_{jk}=U^{i_2}_{jk}$, for any $k_1\neq k_2$, $j\in \{1,2\}$, and $k\in\{0,1,\dots,n-1\}$. In this case, the game is a symmetric game and was widely analyzed recently, such as public goods games and multi-player snowdrift games \cite{govaert2020zero,liang2015analysis}. In public goods games, take $1$-selector as the cooperater and $2$-selector as the defector. In this case, each cooperator contributes an amount $c>0$, and its contribution to a public good is multiplied by an enhancement factor $r$. Each player gets an equal share of the public good. Thus, the cooperator's utility is $U^i_{1,k}=\frac{rc(k+1)}{n}-c$, and the defector's utility is $U^i_{2,j}=\frac{rck}{n}$, for $i\in\{1,\dots,n\}$ and $k\in\{0,1,\dots,n-1\}$. Besides, in multi-player snowdrift games, cooperators need to clear out a snowdrift so that everyone can go on their merry way. All cooperators share a cost $c$ to clear out the snowdrift together, and each one gets a fixed benefit $b$. Thus, the cooperator's utility is $U^i_{1,k}=b-\frac{c}{k+1}$. The defector's utility is $U^i_{2,k}=b$,$k\neq 0$, if there is at least one cooperator, $U^i_{2,0}=0$, since no one clear out the snowdrift.
\end{example}

\begin{example}
In asymmetric public goods game \cite{nockur2021different,reeves2017asymmetric}, players may not get an equal share of the public good due to asymmetry in the distribution of resources. Take $a_i$ as player $i$'s  share factor, where $\sum\limits_i^{n}a_i=1$. Then, player $i$ gets the utilty $U^i_{1,k}=a_irc(k+1)-c$ if it cooperates, while player $i$ gets the utility $U^i_{2,k}=a_irck$ if it defects, for $i\in\{1,\dots,n\}$, and $k\in\{0,1,\dots,n-1\}$. Besides, in asymmetric snowdrift games \cite{du2009asymmetric},  players may have different benefits $b_i$. Thus, the cooperator's utility is $U^i_{1,k}=b_i-\frac{c}{k+1}$. The defector's utility is $U^i_{2,k}=b_i$,$k\neq 0$, if there is at least one cooperator, $U^i_{2,0}=0$, since no one clears out the snodrift.  Moreover, in asymmetric security games, there are two typical players, defenders and attackers. For convenience, consider player $1$ as the defender, and other players are attackers. Then the defender hopes to select the target that faces many attacks, while the attacker has the opponent's preference as the defender. Thus, for player  $1$, i.e. the defender, $U^1_{1,i}>U^1_{1,j}$, $U^1_{2,i}<U^1_{2,j}$, for $i<j$. For $k\neq 1$,  player $k$, i.e. the attacker, $U^1_{1,i}<U^1_{1,j}$, $U^1_{2,i}>U^1_{2,j}$, for $i<j$. 

\end{example}

Besides, denote $\mathcal{S}=\{(1,\dots,1),(1,\dots,1),\dots,(2,\dots,2)\}$ as the set of states, which is composed by the previous actions.  
 $P:\mathcal{S}\times \mathcal{S\times\mathcal{A}}\to[0,1]$ is the transition function, where $P(s'|s,\mathbf{a})$ shows the probability  to the next state $s'\in\mathcal{S}$ from the current state $s$ when players take $\mathbf{a}$, and $\sum\limits_{s'\in\mathcal{S}}P(s'|s,\mathbf{a})=1$, for $s\in\mathcal{S},\mathbf{a}=(a^1,\dots,a^n)\in\mathcal{A}$. Then  $P(s'|s,a^1,\dots,a^n)=1$ if and only if $s'=(a^1,\dots,a^n)$ for any $s\in\mathcal{S}$. Thus, the next state depends on players' strategies and the current state. Each player's strategy depends on the current state. The strategy of player $i$ is a probability distribution $\pi_i$, where $\pi_i(a^i|s)\in\Delta\mathcal{A}_i$ with $\Delta\mathcal{A}_i$ denoting a probability simplex defined on the space $\mathcal{A}_i$. $\pi_i$ is actually a memory-one strategy in the repeated game.
Thus, set $M=\{P(s,s')\}_{s,s'\in\mathcal{S}}$  as the state transition matrix, where $P(s,s')=\pi_1(a^1|s)\pi_2(a^2|s)\dots\pi_n(a^n|s)$, and $s'=(a^1,\dots,a^n)$. Take $a_t^i$ as player $i$'s action in stage $t$. Then player $i$'s utility in stage $t$ is $r_i(a_t^1,a_t^2,\dots,a_t^n)=U^i_{a_t^i,z}$, where $z$ is the number of $a_t^j=1$, $j\neq i$.

The expected long-term discounted  utility of player $i$ in $\mathcal{G}$ is
$$
U_i(\delta,\pi_1,\pi_2,\dots,\pi_n)=\mathbb{E}\left(\lim_{T\to\infty}(1-\delta)\sum\limits_{t=0}^T\delta^Tr_i(a_t^1,a_t^2,\dots,a_t^n)\right), i\in\{ 1,\dots,n\}, 
$$
where  $\left\{a_{t}^i\sim \pi_i(\cdot|s^i_{t})\right\}_{t\geqslant 1}$, and $s_t^i=(a_{t-1}^1,\dots,a_{t-1}^n)$ describe the evolution of states and actions over stage. $\pi_i^0(1)$ denotes the probability of player $i$ choosing action $1$ in stage $0$.  Each player aims to maximize its own expected utility. 

\subsection{Strong Stackelberg
equilibrium strategy}
%We consider two typical strategies in the repeated game, the the strong Stackelberg equilibrium (SSE) strategy and the zero-determinant (ZD) strategy.

 The strong Stackelberg
equilibrium (SSE) strategy can be regarded as the optimal solution for the leader in an actual Stackelberg game. We consider that player $1$ is a leader and declares a strategy in advance, while other players are followers and choose their strategies after observing player $1$'s strategy. The best response (BR) strategy set of player $i$ is denoted as
 $ \textbf{{BR}}_i(\pi_{-i})=\mathop{\text{argmax}}\limits_{\pi_i\in\Delta \mathcal{A}_i} U_i(\delta,\pi_i,\pi_{-i})$. %If player $1$ chooses strategy $\pi_1$, then the set of others' BR strategies is denoted by  $ \textbf{{BR}}(\pi_1)=\mathop{\text{argmax}}\limits_{\pi_i\in\Delta \mathcal{A}_i} U_i(\delta,\pi_i,\pi_{-i})$ 
 Without loss of
generality, followers break ties optimally for the leader
if there are multiple options. In this case, followers choose the following BR strategy profile:
 $$\pi^{\overline{BR}}(\pi_1)\in\overline{\textbf{{BR}}}(\pi_1)=\mathop{{argmax}}\limits_{ \pi_i\in {\textbf{\emph{BR}}}_i({\pi_{-i}}),i\neq 1}U_1(\delta,\pi_1,\pi_2,\dots,\pi_n).$$ 
Player $1$ aims to maximize
its expected utility via the above optimization, and the  equilibrium is defined as 
 below \cite{9722864,lopez2022stationary}.

\begin{definition}\label{def::SSE}
A strategy profile $(\pi_1^{SSE},\pi_2^{SSE},\dots,\pi_n^{SSE})$ is said to be an SSE of $\mathcal{G}$ if
$$
\begin{aligned}
&(\pi_1^{SSE}, \pi_2^{SSE},\dots,\pi_n^{SSE})\in\mathop{{argmax}}\limits_{\pi_1\in\Delta \mathcal{A}_1, \pi_i\in {\textbf{\emph{BR}}}_i({\pi_{-i}}),i\neq 1}U_1(\delta,\pi_1,\pi_2,\dots,\pi_n).
\end{aligned}
$$
\end{definition}
%When player 2 chooses the BR strategy after observing player $1$'s strategy, the SSE strategy $\pi_d^{SSE}$ is optimal for player $1$, and player $1$ has an advantage in guiding the attacker's strategy decision. 

Typically, in the one-to-one situation between player $1$ and player $2$, player $2$ is a follower and chooses the following BR strategy:
 $\pi_2^{\overline{BR}}(\pi_1)\in\overline{\textbf{{BR}}}(\pi_1)=\mathop{\text{argmax}}\limits_{\pi_2\in\textbf{{BR}}_2(\pi_1)} U_1(\delta,\pi_1,\pi_2).$ Different from the multi-player situation, player $2$ does not need to take other followers' actions and preferences into account, when adopting its own strategy. \iffalse
Player $1$ aims to maximize
its expected utility via the above optimization, and the  equilibrium is defined as 
 below \cite{9722864,lopez2022stationary}.
A strategy profile $(\pi_1^{SSE},\pi_2^{SSE})$ is said to be an SSE of $\mathcal{G}$ if
$$
\begin{aligned}
&(\pi_1^{SSE}, \pi_2^{SSE})\in\mathop{\emph{argmax}}\limits_{\pi_1, \pi_2\in {\textbf{\emph{BR}}}({\pi_1})}U_1(\delta,\pi_1,\pi_2).
\end{aligned}
$$
\fi

However, the computation of the SSE in such a repeated game is complex, which is always transformed into a mixed-integer non-linear program with a non-convex optimization objective. The non-convexity makes the computation much difficult. Also, even mixed-integer polynomial programming programs have exponential complexity in the number of states.
Thus, although the SSE strategy is optimal for player $1$, we hope to replace the SSE strategy with another effective strategy, to avoid too much cost of time.

\subsection{Zero-determinant strategy}

Fortunately, methods based on zero-determinant (ZD) strategies have emerged in these years. When adopting a ZD strategy,  one player can unilaterally enforce the two players' expected utilities subjected to a linear relation \cite{press2012iterated}. %It can be regarded as open-loop control, which does not need feedback from the opponent's strategy.
%, which have been widely studied to promote cooperation or unilaterally extortion in public goods game (PGG), human-computer interaction (HCI), and evolutionary games \cite{wang2016extortion,hilbe2013evolution,govaert2020zero}.
%Press and Dyson \cite{press2012iterated} proposed zero-determinant (ZD) strategies, where the player equipped with ZD strategies can unilaterally enforce the two players' expected utilities subjected to a linear relation. Afterward, various ZD strategies were widely studied to promote cooperation or unilaterally extortion in public goods games (PGG), human-computer interaction (HCI), and evolutionary games \cite{wang2016extortion,hilbe2013evolution,govaert2020zero}.
Recently, ZD strategies have been widely studied to promote cooperation or unilaterally extortion in public goods games (PGG), human-computer interaction (HCI), and evolutionary games \cite{cheng2022misperception}.
For this discounted repeated asymmetric game with multiple players $\mathcal{G}$, the player 1's ZD strategy is defined as follows \cite{govaert2020zero}:

\iffalse
Recently, ZD strategies in multi-player games have been widely studied in many situations.
For this discounted repeated asymmetric game $\mathcal{G}$, player $1$'s equalizer ZD strategy is defined as follows :

\fi

\begin{definition}\label{df::ZD}
The strategy $\pi_1^{ZD}$ is called a ZD strategy of player $1$, if there exit constants $\eta,\gamma,\phi$, and weights $\omega_j$ such that
\begin{equation}\label{eq::ZD-multi-definitoin}
\begin{aligned}
\delta\pi_1^{ZD}(1)&= \phi\left(\eta \mathbf{S}^1 -\sum\limits_{j\neq 1}\omega_j\mathbf{S}^j +\gamma \right) -(1-\delta)\pi_1^0(1) \mathbf{1}+\hat{\pi},\\
\pi_1^{ZD}(2)&=1-\pi^{ZD}_1(1),
\end{aligned}
\end{equation}
where $\gamma\in\mathbb{R}$, $\sum\limits_{j\neq 1}\omega_j=1$, $ \hat{\pi}=[1,1,\dots,1,0,\dots,0]^T$ is the repeated strategy, and 
$$
\begin{aligned}
\pi_1(k)&=[\pi_1(k|1,\dots,1),\pi_1(k|1,\dots,2),\dots,\pi_1(k|2,\dots,2)]^T, k\in N,\\
\mathbf{S}^j&=[r(1,\dots,1),r(1,\dots,2),\dots,r(2,\dots,2)]^T.
\end{aligned}
$$

\end{definition}

It is called the zero-determinant strategy because players' expected utilities are subjected to a linear relation: $$ \eta U_1(\delta,\pi_1,\dots,\pi_n)-\omega_2 U_2(\delta,\pi_1,\dots,\pi_n)-\dots-\omega_n U_n(\delta,\pi_1,\dots,\pi_n) +\gamma=0, \text{ for all }\pi_i\in \Delta\mathcal{A}_i,i\neq 1.$$ Further, by taking $\eta=0$, the corresponding strategy is called an equalizer ZD strategy, which can unilaterally control the sum of the opponent's utility \begin{equation}\label{eq::multi-player} \sum\limits_{j\neq 1}\omega_jU_j(\delta,\pi_1,\dots,\pi_n)=\gamma, \text{ for all }\pi_i\in \Delta\mathcal{A}_i,i\neq 1.\end{equation} %The equalizer ZD strategy can unilaterally set the opponent's utility as $ U_2(\pi_1,\pi_2)=-\frac{\gamma}{\beta}$, when $\beta\neq 0$. 
Take $\Xi (\delta)$ as the set of all feasible equalizer strategies. 

Typically, in the in the one-to-one situation between player 1 and player 2, the strategy $\pi_1^{ZD}$ is called a ZD strategy if
$$
\begin{aligned}
\delta\pi_1^{ZD}(1)&=\eta \mathbf{S}^{1} +\beta \mathbf{S}^2 +(\gamma-(1-\delta)\pi_1^0(1)) \mathbf{1}_4+\hat{\pi},\\
\pi_1^{ZD}(2)&=1-\pi^{ZD}_1(1),
\end{aligned}
$$
where $\eta,\beta,\gamma\in\mathbb{R}$, $\hat{\pi}=[1,1,0,0]^T$, and $\pi_1(k)=[\pi_1(k|1,1),\pi_1(k|1,2),\pi_1(k|2,1),\pi_1(k|2,2)]^T,$
$k\in N$. Further, by taking $\eta=0$, the corresponding strategy is called an equalizer ZD strategy, which can unilaterally control the opponent's utility \begin{equation}\label{eq::2-player}\beta U_2(\delta,\pi_1,\pi_2)+\gamma=0, \text{ for all }\pi_2\in \Delta\mathcal{A}_2.\end{equation} The equalizer ZD strategy can unilaterally set the opponent's utility as $ U_2(\pi_1,\pi_2)=-\frac{\gamma}{\beta}$, when $\beta\neq 0$. Equation (\ref{eq::2-player}) is the same as equation (\ref{eq::multi-player}), when $n=2$ and $\omega_2=-\beta$.

Recall that the computation of an SSE strategy has exponential complexity in the number of states. According to \cite{basu2023complexity}, there may be a branch-and-bound algorithm that proves the solution's validity and may take more than $O(2^m)$ iterations, where $m$ is the number of states $s$. Moreover, if we have a linear relation and the corresponding parameters $\omega_j,\gamma$, we only need to solve the parameter $\theta$ which makes the ZD strategy feasible. Then we can directly get the ZD strategy according to equation (\ref{eq::ZD-multi-definitoin}). Actually, it at most spends $O(m)$ times for us to solve $\theta$, since we only need to verify $0\leqslant\pi_1^{ZD}(1)\leqslant 1$. Thus, to reduce the time cost, we aim to find equalizer ZD strategies with acceptable performance in discounted repeated asymmetric games, compared with the baseline SSE strategy.

However, the analysis of a general multi-player situation is more difficult than that in its typical case, the one-to-one situation. For example, the utility functions of players in a multi-player situation are more complex than the utility functions of players in a one-to-one situation obviously. The complex functions make it difficult to compare expected utilities and calculate SSE strategies since it has exponential complexity in the number of states. Secondly, the interaction between followers needs to be considered in multi-player games, while one-to-one does not since it only has one follower. Followers have many internal interactions, such as cooperation, competition, and so on, in multi-player games. In the one-to-one game, player $2$ only needs to maximize its own utility. Therefore, we hope to consider the one-to-one situation in the next section.

\section{One-to-one situation}
As a fundamental case of a multi-player situation, we consider one-to-one situations in this section. Actually, the two-player game is one of the most popular game mechanisms, where many games are based on the discussion of two-person games. For example, a general security game is based on a typical model with one attacker and one defender. Whether two players cooperate has also inspired the analysis of snowdrift games and public goods games. Moreover, the two-player asymmetric game in this paper has also been widely analyzed. For example, in an asymmetric public goods game between two players, players may not get an equal share of the public good due to asymmetry in the distribution of resources \cite{zhu2014promotion}. Besides, In asymmetric snowdrift games, one player may gain more than the other \cite{han2023complex}. Also, in security games, the defender and the attacker have different preferences, where the defender tends to protect the vulnerable target and the attacker tends to implement invasions on the unprotected target \cite{cheng2023zero}.

%\subsection{Performance}

First, we need to verify the existence for equalizer ZD strategies in the discounted repeated asymmetric game, since not all linear relations can be enforced by feasible ZD strategies.
%Since player $1$'s equalizer ZD strategy must belong to the implementer's strategy set, there may not exist an equalizer ZD strategy. Thus, the existence of an equalizer ZD strategy is well worth exploring in the discounted repeated asymmetric game.  
For convenience, take the set$$\begin{aligned}\Lambda(\delta)=
\{\phi:& \phi \leqslant \frac{\min\limits_{x\in\{(1,1),(1,0)\},y\in\{(2,1),(2,0)\}}\{U_{x}^2-U_y^2\}}{1-\delta},\\
&\phi\geqslant\max\{\frac{|U_{1,1}^2\!-\!U_{1,0}^2|}{\delta},\!\frac{|U_{2,1}^2\!-\!U_{2,0}^2|}{\delta},\!\frac{\max\limits_{x\in\{(1,1),\!(1,0)\},y\in\{(2,1),\!(2,0)\}}\{U_{x}^2\!-\!U_y^2\}}{1+\delta}\}\}
.
\end{aligned}$$
Actually, when we verify whether $\phi$ is in the set $\Lambda(\delta)$, we only need to check $\phi$ with ten compared elements. 
The following lemma provides a sufficient condition for the existence of equalizer ZD strategies.

\begin{lemma}\label{th::ZD-1}

An equalizer ZD strategy exists if  $\Lambda(\delta)$ is nonempty.
\end{lemma}

Lemma \ref{th::ZD-1} shows that the linear relation between players' expected utility can be enforced by player $1$'s equalizer ZD strategy if $\Lambda(\delta)$ is nonempty. %We present a simplified existence condition by comparing values of several elements as shown in $\Lambda(\delta)$.   %Moreover, the condition in Theorem \ref{th::1} is also a necessary and sufficient condition for the existence of ZD strategy in repeated security games with two targets.

Next, the equalizer ZD strategy of player $1$ can unilaterally control the opponent's utility, but the performance of this unilateral control is indeed bounded. Thus, we need to explore bounds of the opponent utility when adopting equalizer ZD strategies. Denote
$$\begin{aligned}\Gamma^{+}_1(\delta)=\max\limits_{\phi\in\Lambda(\delta)} \min \left\{ U_x^2-\frac{(1-\delta)(1-\pi_1^0(1))}{\phi}, U_y^2+\frac{\delta+(1-\delta)\pi_1^0(1)}{\phi} ,\right. \\
\left.x \in \{(1,1),(1,0)\},y\in\{(2,1),(2,0)\}\right\},\end{aligned}$$ and 
$$\begin{aligned}\Gamma^{-}_1(\delta)=\min\limits_{\phi\in\Lambda(\delta)} \max \left\{ U_x^2-\frac{1-(1-\delta)\pi_1^0(1)}{\phi}, U_y^2+\frac{(1-\delta)\pi_1^0(1)}{\phi} ,\right. \\
\left.x \in \{(1,1),(1,0)\},y\in\{(2,1),(2,0)\}\right\}.\end{aligned}$$  Then we give a bound of the opponent's utility when adopting equalizer ZD strategies in the following result.

\begin{theorem}\label{thh::ZD-1} If  $\Lambda(\delta)$ is nonempty, then there exists an equalizer ZD strategy such that $  U_2= \gamma$ with
$\Gamma^{-}_1(\delta)\leqslant\gamma\leqslant \Gamma^{+}_1(\delta).$
\end{theorem}

\begin{myproof}
Since $\Lambda(\delta)$ is nonempty, there are $\beta,\gamma,$ and $\phi$, such that
$$
\begin{aligned}
(1-\delta)(1-\pi_1^0(1))\leqslant -\phi( \beta U_{1,1}^2 + \gamma ) \leqslant 1-(1-\delta)\pi_1^0(1),\\
(1-\delta)\pi_1^0(1)\leqslant \phi(\beta U_{2,1}^2 + \gamma ) \leqslant \delta+(1-\delta)\pi_1^0(1),\\
(1-\delta)(1-\pi_1^0(1))\leqslant -\phi( \beta U_{1,0}^2 + \gamma )\leqslant 1-(1-\delta)\pi_1^0(1),\\
(1-\delta)\pi_1^0(1)\leqslant \phi( \beta U_{2,0}^2 + \gamma ) \leqslant \delta +(1-\delta)\pi_1^0(1).
\end{aligned}
$$
Without loss of generality, take $\beta=-1$. Then
\begin{subequations}
\begin{equation}\label{eq::2a} (1-\delta)(1-\pi_1^0(1))\leqslant -\phi( -U_{1,1}^2 + \gamma ) \leqslant 1-(1-\delta)\pi_1^0(1) , \end{equation}
\begin{equation}\label{eq::2b} (1-\delta)(1-\pi_1^0(1))\leqslant -\phi( -U_{1,0}^2 + \gamma )\leqslant 1-(1-\delta)\pi_1^0(1),\end{equation}
\begin{equation}\label{eq::2c}(1-\delta)\pi_1^0(1)\leqslant \phi( -U_{2,1}^2 + \gamma ) \leqslant \delta+(1-\delta)\pi_1^0(1),\end{equation}
\begin{equation}\label{eq::2d}(1-\delta)\pi_1^0(1)\leqslant \phi( -U_{2,0}^2 + \gamma ) \leqslant \delta +(1-\delta)\pi_1^0(1).\end{equation}
\end{subequations}

First, suppose $\phi>0$. By multiplying both sides of (\ref{eq::2a}) (\ref{eq::2b}) by $-\frac{1}{\phi}$, and (\ref{eq::2c}) (\ref{eq::2d}) by $\frac{1}{\phi}$, the above inequalities can be converted to
$$
\begin{aligned}
-\frac{1-(1-\delta)\pi_1^0(1)}{\phi} \leqslant -U_{1,1}^2 + \gamma  \leqslant - \frac{(1-\delta)(1-\pi_1^0(1))}{\phi},\\
-\frac{1-(1-\delta)\pi_1^0(1)}{\phi}\leqslant - U_{1,0}^2 + \gamma \leqslant -\frac{(1-\delta)(1-\pi_1^0(1))}{\phi},\\
\frac{(1-\delta)\pi_1^0(1)}{\phi}\leqslant  -U_{2,1}^2 + \gamma  \leqslant \frac{\delta+(1-\delta)\pi_1^0(1)}{\phi},\\
\frac{(1-\delta)\pi_1^0(1)}{\phi}\leqslant  -U_{2,0}^2 + \gamma\leqslant \frac{\delta +(1-\delta)\pi_1^0(1)}{\phi}.
\end{aligned}
$$
Therefore, we have following inequalities:
$$\gamma\leqslant\min \{ U_x^2-\frac{(1-\delta)(1-\pi_1^0(1))}{\phi}, U_y^2+\frac{\delta+(1-\delta)\pi_1^0(1)}{\phi} ,x \in \{(1,1),(1,0)\},y\in\{(2,1),(2,0)\}\},$$ and 
$$\gamma\geqslant
 \max \{ U_x^2-\frac{1-(1-\delta)\pi_1^0(1)}{\phi}, U_y^2+\frac{(1-\delta)\pi_1^0(1)}{\phi} ,x \in \{(1,1),(1,0)\},y\in\{(2,1),(2,0)\}\}.$$
Since $\phi\in\Lambda(\delta)$,  $\Gamma^{-}_1(\delta)\leqslant\gamma\leqslant \Gamma^{+}_1(\delta).$ Moreover, we can get the same result considering the case $\phi<0$.
\end{myproof}

We learn from Theorem  \ref{thh::ZD-1} that player $1$ can unilaterally set the opponent's utility in $[\Gamma^{-}_1(\delta),\Gamma^{+}_1(\delta)]$ and needs not care about what strategies the opponent selects. Actually, for any $\gamma \in [\Gamma^{-}_1(\delta),\Gamma^{-}_1(\delta)]$, an equalizer ZD strategy is feasible for the player $1$ to enforce the linear relation.  Thus, it gives player $1$ a choice, a ZD strategy, to enforce unilateral control to the opponent's utility in the repeated game $\mathcal{G}$. 

\iffalse
The proof of Theorem \ref{thh::ZD-1} is mainly divided into the following steps.  At first, we analyze that player is able to set the opponent's utility as $\Gamma^{+}_1(\delta)$ $(\Gamma^{-}_1(\delta))$ when $\Lambda(\delta)$ is nonempty.  Besides, we prove that $\Gamma^{+}_1(\delta)$ $(\Gamma^{-}_1(\delta))$ is the upper (lower) bound of the opponent's utility when the player $1$ chooses equalizer ZD strategies. Finally, we show the corresponding equalizer ZD strategy is feasible for any $\gamma \in [\Gamma^{-}_1(\delta),\Gamma^{-}_1(\delta)]$. 
We learn from Theorem  \ref{thh::ZD-1} that, player $1$ can unilaterally set the opponent's utility in $[\Gamma^{-}_1(\delta),\Gamma^{+}_1(\delta)]$, and does not care about what strategies the opponent selects.  Thus, it gives player $1$ a choice, ZD strategy to enforce unilateral control to the opponent's utility in the repeated game $\mathcal{G}$. 
\fi

Next, we analyze how player $1$ gains the best utility by the equalizer ZD strategy and reduces the utility gap between SSE strategies and ZD strategies. 
According to Definition \ref{def::SSE}, the SSE strategy always brings the highest utility for the leader with the follower's BR strategy. Thus, we need to show the upper limit of the performance of ZD strategies, and the loss compared with the SSE strategy. Actually,  the discounted repeated game has a complex expected utility $ U_2(\pi_1,\pi_2)$, which can be written as the division of two determinants in the repeated game with average-sum utilities \cite{press2012iterated}. For convenience, we denote $\pi_1^{+}(\pi_1^{-})$ as the equalizer ZD strategy to enforce $U_2=\Gamma_1^{+}(\delta)(\Gamma_1^{-}(\delta))$, and $U_1^{SSE}$ as player $1$'s utility in SSE.  Denote $\Pi(\mathcal{B})=\{\pi_2\in\Delta\mathcal{B}|\pi_2(b|s)\in\{0,1\},\forall b,s\}$, and
$$\begin{aligned}&D(\delta,\pi_1,\pi_2,\mathbf{f})=\left[\begin{array}{llll}
\delta\pi_1(1|1,1)\pi_2(1|1,1)\!-\!1  & \delta\pi_1(1|1,1)\!-\!1 &\delta\pi_2(1|1,1) \!-\!1 &f_1\\
\delta\pi_1(1|1,2)\pi_2(1|1,2) & \delta\pi_1(1|1,2)\!-\!1 & \delta\pi_2(1|1,2)&f_2 \\
\delta\pi_1(1|2,1)\pi_2(1|2,1) & \delta\pi_1(1|2,1)& \delta \pi_2(1|2,1)\!-\!1&f_3\\
\delta\pi_1(1|2,2)\pi_2(1|2,2) & \delta\pi_1(1|2,2) &\delta\pi_2(1|2,2) & f_4
\end{array}\right],
\end{aligned}$$ %Thus, $\Pi(\mathcal{B})$ has $16$ elements. 
\begin{equation}\label{eq::Ldelta}L(\delta)=\max\limits_{\pi_1\in\{\pi_1^{+},\pi_1^{-}\},\pi_2\in \Pi(\mathcal{B})} \frac{(1-\delta)D(\delta,\pi_1,\pi_2,\mathbf{S}^1)}{det(\mathbf{I}_4-\delta M)}.\end{equation}
Then we show the utility gap of player $1$'s utilities between an equalizer ZD strategy and the SSE strategy in the following theorem.

\begin{theorem}\label{th::ZD-SSE-1}
If $\Lambda(\delta)$ is nonempty and $\pi_1^0(1)=\pi_2(0)=0$, then we get a utility bound below
$$\begin{aligned}&\min\limits_{\pi_1^{ZD}\in \Xi (\delta)}U_1\left(\delta,\pi_1^{SSE}, \pi_2^{\overline{BR}}(\pi_1^{SSE})\right)\!-\!U_1\left(\delta,\pi_1^{ZD}, \pi_2^{\overline{BR}}(\pi_1^{ZD})\right) 
=
U_1^{SSE}-L(\delta).
\end{aligned}$$
\end{theorem}
\begin{myproof}
For any $\pi_1,\pi_2$, take $v(0)=[ \pi_1^0(1)\pi_2^0(1),\pi_1^0(1)(1-\pi_2^0(1)),(1-\pi_1^0(1))\pi_2^0(1),(1-\pi_1^0(1))(1-\pi_2^0(1))]$ as the probability of the initial state, and the state transition matrix as $$\begin{aligned}&M=\left[\begin{array}{llll}
\pi_1(1|1,1)\pi_2(1|1,1) & \pi_1(1|1,1)\pi_2(2|1,1) &\pi_1(2|1,1)\pi_2(1|1,1)&\pi_1(2|1,1)\pi_2(2|1,1)\\
\pi_1(1|1,2)\pi_2(1|1,2) & \pi_1(1|1,2)\pi_2(2|1,2) &\pi_1(2|1,2)\pi_2(1|1,2)&\pi_1(2|1,2)\pi_2(2|1,2)\\
\pi_1(1|2,1)\pi_2(1|2,1) & \pi_1(1|2,1)\pi_2(2|2,1) &\pi_1(2|2,1)\pi_2(1|2,1)&\pi_1(2|2,1)\pi_2(2|2,1)\\
\pi_1(1|2,2)\pi_2(1|2,2) & \pi_1(1|2,2)\pi_2(2|2,2) &\pi_1(2|2,2)\pi_2(1|2,2)&\pi_1(2|2,2)\pi_2(2|2,2)\\
\end{array}\right]
\end{aligned}.$$ Then player $1$'s expected long-term discounted utility can be written as
$$\begin{aligned}
&U_1(\delta,\pi_1,\pi_2)\\
=&(1-\delta)\sum\limits_{t=0}^{\infty}\delta^tv(0)M^t\cdot \mathbf{S}^1\\
=&(1-\delta)v(0)\left(I_4-\delta M\right)^{-1}\cdot \mathbf{S}^1\\
=&\frac{1-\delta}{det(I-\delta M)}v(0)\left[\begin{array}{c}
A_{11}U_{11}^1+A_{21}U_{12}^1+A_{31}U_{21}^1+A_{41}U_{22}^1\\
A_{12}U_{11}^1+A_{22}U_{12}^1+A_{32}U_{21}^1+A_{42}U_{22}^1\\
A_{13}U_{11}^1+A_{23}U_{12}^1+A_{33}U_{21}^1+A_{43}U_{22}^1\\
A_{14}U_{11}^1+A_{24}U_{12}^1+A_{34}U_{21}^1+A_{44}U_{22}^1\\
\end{array} \right],
\end{aligned}$$
where $A_{ij}$ is the element of the adjugate matrix of $\delta M$.
Denote $det[M,\mathbf{f},i]$ as the determinate of the matrix where the $i$th column of $\delta M-\mathbf{I}$ is replaced by $\mathbf{f}$. For example,
$$\begin{aligned}&det[ M,\mathbf{f},4]=\\&\left[\begin{array}{llll}
\delta\pi_1(1|1,1)\pi_2(1|1,1)-1 & \delta\pi_1(1|1,1)\pi_2(2|1,1) &\delta\pi_1(2|1,1)\pi_2(1|1,1)&f_1\\
\delta\pi_1(1|1,2)\pi_2(1|1,2) & \delta\pi_1(1|1,2)\pi_2(2|1,2)-1 &\delta\pi_1(2|1,2)\pi_2(1|1,2)&f_2\\
\delta\pi_1(1|2,1)\pi_2(1|2,1) &\delta \pi_1(1|2,1)\pi_2(2|2,1) &\delta\pi_1(2|2,1)\pi_2(1|2,1)-1&f_3\\
\delta\pi_1(1|2,2)\pi_2(1|2,2) &\delta \pi_1(1|2,2)\pi_2(2|2,2) &\delta\pi_1(2|2,2)\pi_2(1|2,2)&f_4\\
\end{array}\right]
\end{aligned}.$$
Thus, player $1$'s expected utility can be written as:
$$\begin{aligned}
&U_1(\delta,\pi_1,\pi_2)\\
=&\frac{1-\delta}{det(I-\delta M)}v(0)\left[\begin{array}{c}
det[M,\mathbf{S}^1,1]\\
det[M,\mathbf{S}^1,2]\\
det[M,\mathbf{S}^1,3]\\
det[M,\mathbf{S}^1,4]
\end{array} \right]\\
=&\frac{1-\delta}{det(I-\delta M)}( \pi_1^0(1)\pi_2^0(1)det[M,\mathbf{S}^1,1]+\pi_1^0(1)(1-\pi_2^0(1))det[M,\mathbf{S}^1,2]\\
&\quad\quad\quad\quad\quad\quad+(1-\pi_1^0(1))\pi_2^0(1)det[M,\mathbf{S}^1,3]+(1-\pi_1^0(1))(1-\pi_2^0(1))det[M,\mathbf{S}^1,4])
\end{aligned}$$
Since $\pi_1^0(1)=\pi_2^0(1)=0$, we have $U_1(\delta,\pi_1,\pi_2)=\frac{(1-\delta)det[M,\mathbf{S}^1,4]}{det(I-\delta M)}= \frac{(1-\delta)D(\delta,\pi_1,\pi_2,\mathbf{S}^1)}{det(\mathbf{1}_4-\delta M)}.$

According to Theorem \ref{thh::ZD-1}, player $1$ can unilerally set the opponents' utility as  $  U_2= \gamma$, where $\Gamma^{-}_1(\delta)\leqslant\gamma\leqslant \Gamma^{+}_1(\delta).$ Notice that $ U^2_y\leqslant \Gamma^{-}_1(\delta),\Gamma^{+}_1(\delta)\leqslant U^2_{x}$, for $x \in \{(1,1),(1,0)\},y\in\{(2,1),(2,0)\}$.
 We first consider $$\max\{U^1_{x},x \in \{(1,1),(1,0)\}\}>\max\{U^1_y,y\in\{(2,1),(2,0)\}\}.$$ Then player $1$'s best equalizer ZD strategy is $\pi_1^+$, and $\pi_2^{\overline{BR}}(\pi_1)\in \Pi(\mathcal{B})$. Thus, $U_1(\delta,\pi_1^+,\pi_2^{\overline{BR}}(\pi_1^+))= \max\limits_{\pi_2\in \Pi(\mathcal{B})} \frac{(1-\delta)D(\delta,\pi_1^+,\pi_2,\mathbf{S}^1)}{det(\mathbf{1}_4-\delta M)}.$ Similarly, when $$\max\{U^1_{x}\!,x \in \{(1,1),(1,0)\}\}\!<\!\max\{U^1_y,y\in\!\{(2,1),(2,0)\}\!\},$$ we have $U_1(\delta,\pi_1^-,\pi_2^{\overline{BR}}(\pi_1^-))= \max\limits_{\pi_2\in \Pi(\mathcal{B})} \frac{(1-\delta)D(\delta,\pi_1^-,\pi_2,\mathbf{S}^1)}{det(\mathbf{1}_4-\delta M)}$. Thus, $\max\limits_{\pi_1^{ZD}\in \Xi (\delta)} U_1(\delta,\pi_1^{ZD},\pi_2^{\overline{BR}}(\pi_1^{ZD}))=L(\delta).$ Finally, we have 
$$\begin{aligned}\min\limits_{\pi_1^{ZD}\in \Xi (\delta)}U_1\left(\delta,\pi_1^{SSE}, \pi_2^{\overline{BR}}(\pi_1^{SSE})\right)\!-\!U_1\left(\delta,\pi_1^{ZD}, \pi_2^{\overline{BR}}(\pi_1^{ZD})\right)
=
U_1^{SSE}-L(\delta),
\end{aligned}$$which yields the conclusion
\end{myproof}

%$L(\delta)\!=\!\!\!\! \!\!\max\limits_{x\in \{(1,1),(1,0)\},y\in\{(2,1),(2,0)\},Q \in \{\Gamma^{+}_1(\delta),\Gamma^{-}_1(\delta)\} }\!\!\!\!\!\frac{(Q-U_{y}^2)(U_{x}^1-U_{y}^1)}{U_{x}^2-U_y^2}\!+\!U_y^1$.

\iffalse

The proof of Theorem \ref{th::ZD-SSE-1} is mainly divided into the following steps. First, we prove that player $1$'s discounted long-term utility is equal to $\frac{(1-\delta)D(\delta,\pi_1,\pi_2,\mathbf{S}^1)}{det(\mathbf{1}_4-\delta M)}$, when $\pi_1^0(1)=\pi_2(0)=0$. Besides, considering player $1$'s equalizer ZD strategy, the BR strategy set of player $2$ satisfies $\overline{\textbf{{BR}}}(\pi_1)\subseteq\Pi(\mathcal{B})\subseteq{\textbf{{BR}}}(\pi_1)$. Finally, we prove that $L({\delta})$ is the optimal utility for player $1$ with an equalizer ZD strategy. 
\fi

Theorem \ref{th::ZD-SSE-1} shows that the defender can adopt ZD strategies to get a tolerable loss in the utility compared with SSE strategies. 
%Actually, the computation of $L({\delta})$ in (\ref{eq::Ldelta}) is a sample since it only finds the maximum form $32$ values. 
On this basis, when setting the opponent's utility as $\Gamma^{+}(\delta)$ or $\Gamma^{-}(\delta)$,  player $1$ can receive the best utility of adopting equalizer ZD strategies. Notice that computing an SSE strategy is difficult and yields computing resources. If player $1$ can endure the bounded loss, then it can adopt the corresponding equalizer ZD strategy to avoid the cost of computing the SSE strategy. Thus, it also can be regarded as a tradeoff for different strategy choices.
%On the other hand, if the defender can endure the bounded loss, then adopting the corresponding ZD strategy is also a good choice to avoid the complex calculation for SSE strategies since deriving SSE strategies needs to solve a bi-level optimization problem.

\section{Multi-player situation} 
It is time to consider the general case, multi-player situation, and many practical problems involve multiple players' interactions
\cite{zhang2023distributed,govaert2020zero,chen2020learning,mutzari2021coalition}. For example, in security games among multiple players, defenders often need to defend against the intrusion of many attackers of different types \cite{9722864}. Also, the information interaction among drones is also realistic due to UAV networks \cite{sanjab2020game}. In fact, the situation where each player has two actions to choose from is an important multi-player game. For example, in the multi-player public goods game and the multi-player snowdrift game, players can choose to cooperate and work together to complete the task, or refuse to make an effort and only enjoy the benefits \cite{liang2015analysis}. Next, we discuss the multi-player game in which players have two actions.

Similar to Section 3, we first need to examine the conditions under which the ZD policy exists.
It is the foundation when the player has an alternative ZD strategy to replace the SSE strategy. Take
$$\begin{aligned}\Lambda_2(\delta)=\left\{(\phi,\gamma):\right.& \phi(-\sum\limits_{j\neq 1}\omega_j S^j+\gamma)\geqslant(1-\delta)\pi_1^0(1)+\hat{\pi}, \\ &\phi(-\sum\limits_{j\neq 1}\omega_j S^j+\gamma)\leqslant \delta\!+\!(1\!-\!\delta)\pi_1^0(1)\mathbf{1}\!+\!\hat{\pi}\},\end{aligned}$$
$\Gamma^{+}_2(\delta)=\max\limits_{(\phi,\gamma)\in\Lambda_2(\delta)}\gamma$, and 
$\Gamma^{-}_2(\delta)=\min\limits_{(\phi,\gamma)\in\Lambda_2(\delta)}\gamma$. We have the following theorem to show player 1's control on bounds of the weighted sum of the opponents' utilities.

\begin{theorem}\label{cor::ZD-multi}

 If  $\Lambda_2(\delta)$ is nonempty, then there exists an equalizer ZD strategy such that $ \sum\limits_{i\neq 1}\omega_jU_j(\delta,\pi_1,\dots,\pi_n)= \gamma$  with 
$\Gamma^{-}_2(\delta)\leqslant\gamma\leqslant \Gamma^{+}_2(\delta)$.
\end{theorem}

\begin{myproof}
Since $\Lambda_2(\delta)$ is nonempty, there are $\omega_j,\gamma,$ and $ \phi$, such that
$$\begin{aligned}& \phi(-\sum\limits_{j\neq 1}\omega_j S^j+\gamma)\geqslant(1-\delta)\pi_1^0(1)+\hat{\pi}, \\ &\phi(-\sum\limits_{j\neq 1}\omega_j S^j+\gamma)\leqslant \delta\!+\!(1\!-\!\delta)\pi_1^0(1)\mathbf{1}_{2n}\!+\!\hat{\pi},\end{aligned}$$
Therefore, we have following inequalities:
\begin{equation}\label{eq::3.41} (1-\delta)\pi_1^0(1) \mathbf{1}_{2n}+\hat{\pi}\leqslant \phi\left( -\sum\limits_{j\neq 1}\omega_j\mathbf{S}^j +\gamma \right)\leqslant (1-\delta)\pi_1^0(1) \mathbf{1}_{2n}+\hat{\pi}+\delta\mathbf{1}_{2n}.\end{equation}
\iffalse
$$ (1-\delta)\pi_1^0(1) \mathbf{1}_{2n}+\hat{\pi}\leqslant \phi\left( -\sum\limits_{j\neq 1}\omega_j\mathbf{S}^j +\gamma \right)\leqslant (1-\delta)\pi_1^0(1) \mathbf{1}_{2n}+\hat{\pi}+\delta\mathbf{1}_{2n}.$$ 
\fi
Take $\pi_1^{ZD}$ as follows:
$$
\begin{aligned}
\delta\pi_1^{ZD}(1)&= \phi\left( -\sum\limits_{j\neq 1}\omega_j\mathbf{S}^j +\gamma \right) -(1-\delta)\pi_1^0(1) \mathbf{1}_{2n}+\hat{\pi},\\
\pi_1^{ZD}(2)&=1-\pi^{ZD}_1(1),
\end{aligned}
$$
Due to (\ref{eq::3.41}), we have $0\leqslant \phi\left( -\sum\limits_{j\neq 1}\omega_j\mathbf{S}^j +\gamma \right) -(1-\delta)\pi_1^0(1) \mathbf{1}_{2n}+\hat{\pi}\leqslant\delta\mathbf{1}_{2n} $. Thus, $0\leqslant\pi_1^{ZD}(1)\leqslant\mathbf{1}_{2n} $, and  $\pi_1^{ZD}$  is a feasible equalizer ZD strategy for player 1. Moreover, $\min\limits_{(\phi,\gamma)\in\Lambda_2(\delta)}\gamma\leqslant\gamma\leqslant\max\limits_{(\phi,\gamma)\in\Lambda_2(\delta)}\gamma$. Then $\Gamma^{-}_2(\delta)\leqslant\gamma\leqslant \Gamma^{+}_2(\delta)$.
\end{myproof}

 Theorem  \ref{cor::ZD-multi} indicates that player 1 can unilaterally set the sum of opponents' utilities in $[\Gamma^{-}_2(\delta),\Gamma^{+}_2(\delta)]$ and needs not care about what strategies the opponents select. Actually, for any $\gamma \in [\Gamma^{-}_2(\delta),\Gamma^{-}_2(\delta)]$, an equalizer ZD strategy is feasible for the player 1 to enforce the linear relation.  Thus, the player is able to make use of unilateral control in the sum of opponents' utilities to achieve its intention.

\iffalse
Thus, it gives player 1 a choice, ZD strategy to enforce unilateral control to the opponent's utility in the repeated game $\mathcal{G}$.

In order to prove Theorem \ref{cor::ZD-multi}, we first need to  analyze that, for any $\phi,\gamma\in\Gamma_2(\delta)$, $\pi_1(a|s)$ is in the range $[0,1]$ when $\pi_1$ is selected from (\ref{eq::ZD-multi-definitoin}). Then we show the corresponding equalizer ZD strategy is feasible for $\gamma \in [\Gamma^{-}_2(\delta),\Gamma^{-}_2(\delta)]$, which is similar to Theorem \ref{thh::ZD-1}.

\fi
\iffalse
The analysis of Theorem \ref{cor::ZD-multi} is similar to Theorem \ref{thh::ZD-1}. 

 mainly divided into the following steps.  At first, considering $\Gamma_2(\delta)$ is nonempty, we analyze that, for any $\phi,\gamma\in\Gamma_2(\delta)$, $\pi_1(a|s)$ is in the range $[0,1]$ when $\pi_1$ is selected from (\ref{eq::ZD-multi-definitoin}). Besides, we prove that $\Gamma^{+}_2(\delta)(\Gamma^{-}_2(\delta))$ is the upper (lower) bound of the weighted sum of the opponents’ utilities when the player $1$ chooses equalizer ZD strategies, for given $\omega_j,j\neq 1$. Finally, we show the corresponding equalizer ZD strategy is feasible for $\gamma \in [\Gamma^{-}_2(\delta),\Gamma^{-}_2(\delta)]$.

\fi
%Next, inspired by Theorem \ref{th::ZD-SSE-1}, we analyze how player $1$ gets the best utilities through equalizer ZD strategy as reduces the utility loss from adopting the SSE strategy.  

Next, we analyze how player $1$ gains the best utility by the equalizer ZD strategy and reduces the utility gap between SSE strategies and ZD strategies. 
Similar to two-player situations, we need to provide an upper bound of the performance of ZD strategies, and the loss compared with the SSE strategy in multi-player situations. 
Take $\pi_1^{+}$ $(\pi_1^{-})$ as the equalizer ZD strategy to enforces $\sum\limits_{i\neq 1}\omega_jU_j(\delta,\pi_1,\dots,\pi_n)=\Gamma_2^{+}(\delta)$ $(\Gamma_2^{-}(\delta))$.  Let $\Pi_{-1}(\mathcal{B})=\{\pi_j\in\Delta\mathcal{A}|\pi_j(a|s)\in\{0,1\},\forall a,s,j\neq 1\}$. Denote  $s=[s(1),s(2),\dots, s(m)]=[(1,1,\dots,1),(1,1,\dots,2),\dots, (2,2,\dots,2)]^T$ as all possible state, where $m=2^n$.  Then take $D(\delta,\pi_1,\dots,\pi_{n},\mathbf{f})$ as an extension of $D(\delta,\pi_1,\pi_2,\mathbf{f})$, where
$$\begin{aligned}&D(\delta,\pi_1,\dots,\pi_{n},\mathbf{f})=\\&\left[\begin{array}{lllll}
\delta\pi(s(1)|s(1))\!-\!1  & \delta\pi(s(2)|s(1))\!-\!1 &\delta\pi(s(3)|s(1)) \!-\!1 &\dots&f_1\\
\delta\pi(s(1)|s(2)) & \delta\pi(s(2)|s(2))\!-\!1 & \delta\pi(s(3)|s(2))&\dots&f_2 \\
\vdots\\
\delta\pi(s(1)|s(m)) & \delta\pi(s(2)|s(m)) &\delta\pi(s(3)|s(m))&\dots & f_m
\end{array}\right],
\end{aligned}$$  and $\pi(s(i)|s(j))=\prod\limits_{k=1}^{n}\pi_{k}(g(i,k)|h(j,k))$, $g(i,k)=\left(\left(\left\lceil \frac{i}{2^{n-k}}\right\rceil+1\right)\!\!\!\!\mod2\right)+1$., and $h(j,k)= (g(j,k),\sum\limits_{z\neq k}1_{g(j,z)=1})$. Set \begin{equation}\label{eq::Ldelta_multi}L_2(\delta)=\max\limits_{\pi_1\in\{\pi_1^{+},\pi_1^{-}\},\pi_{-1}\in \Pi_{-1}(\mathcal{B})} \frac{(1-\delta)D(\delta,\pi_1,\dots,\pi_{n},\mathbf{S}^1)}{det(I-\delta M)}.\end{equation}Then we have the following result for an upper bound of player 1's utility in multi-player situations.

\begin{theorem}\label{th::ZD-SSE-multi-compare}
If $\Lambda_2(\delta)$ is nonempty,  $S^{j_1}=S^{j_2}$, and $\pi_i(0)=0$ for $j_1,j_2\neq 1$, $i\in N$, then we get a utility bound below
$$\begin{aligned}&\min\limits_{\pi_1^{ZD}\in \Xi_1 (\delta)}U_1\left(\delta,\pi_1^{SSE}, \pi_{-1}^{\overline{BR}}(\pi_1^{SSE})\right)\!-\!U_1\left(\delta,\pi_1^{ZD}, \pi_{-1}^{\overline{BR}}(\pi_1^{ZD})\right) =
U_1^{SSE}-L_2(\delta).
\end{aligned}$$

\end{theorem}

\begin{myproof}
For any $\pi_1,\pi_2,\dots,\pi_n$, take $v(0)=[ \pi^0(s(1)),\dots,\pi^0(s(m))]$ as the probability of the initial state, and $$\begin{aligned}&M=\left[\begin{array}{cccc}
\pi(s(1)|s(1)) &\pi(s(2)|s(1)) & \dots & \pi(s(m)|s(1))\\
\pi(s(1)|s(2)) &\pi(s(2)|s(2)) & \dots & \pi(s(m)|s(2))\\
\vdots & \vdots& \ddots& \vdots\\
\pi(s(1)|s(m)) &\pi(s(2)|s(m)) & \dots & \pi(s(m)|s(m))
\end{array}\right]
\end{aligned}$$ as the state transition matrix. Then player 1’s expected long-term discounted utilities can
be written as

$$\begin{aligned}
&U_1(\delta,\pi_1,\pi_2,\dots,\pi_n)\\
=&(1-\delta)\sum\limits_{t=0}^{\infty}\delta^tv(0)M^t\cdot \mathbf{S}^1\\
=&(1-\delta)v(0)\left(I_{m}-\delta M\right)^{-1}\cdot \mathbf{S}^1\\
=&\frac{1-\delta}{det(I-\delta M)}v(0)\left[\begin{array}{c}
A_{11}\mathbf{S}^1_1+A_{21}\mathbf{S}^1_2+\dots +A_{m1}\mathbf{S}^1_m\\
A_{12}\mathbf{S}^1_1+A_{22}\mathbf{S}^1_2+\dots +A_{m2}\mathbf{S}^1_m\\
\vdots\\
A_{1m}\mathbf{S}^1_1+A_{2m}\mathbf{S}^1_2+\dots +A_{mm}\mathbf{S}^1_m
\end{array} \right],
\end{aligned}$$
where $A_{ij}$ is the element of the adjugate matrix of $\delta M-\mathbf{I}$.

When $\pi_i^0(1)=0$, for $i=\{1,\dots,n\}$, we have
$$\pi^0(s(i))=\prod\limits_{k=1}^{n}\pi^0_{k}(g(i,k)),$$ where $g(i,k)=\left(\left(\left\lceil \frac{i}{2^{n-k}}\right\rceil+1\right)\!\!\!\!\mod2\right)+1.$ Then
$$\pi^0(s(m))=\prod\limits_{k=1}^{n}\pi_{k}^0(g(m,k))=\prod\limits_{k=1}^{n}\pi_{k}^0(\left(\left(\left\lceil \frac{i}{2^{n-k}}\right\rceil+1\right)\!\!\!\!\mod2\right)+1)=\prod\limits_{k=1}^{n}\pi_{k}^0(2)=1,$$
and, for $j<m$,
$$\pi^0(s(j))=\prod\limits_{k=1}^{n}\pi_{k}^0(g(j,m))=\prod\limits_{k=1}^{n}\pi_{k}^0(\left(\left(\left\lceil \frac{i}{2^{n-k}}\right\rceil+1\right)\!\!\!\!\mod2\right)+1)=0 .$$
Thus, player $1$'s expected utility can be written as:
$$\begin{aligned}
&U_1(\delta,\pi_1,\pi_2,\dots,\pi_n)\\
=&\frac{1-\delta}{det(I-\delta M)}[0,0,\dots,0,1]\left[\begin{array}{c}
A_{11}\mathbf{S}^1_1+A_{21}\mathbf{S}^1_2+\dots +A_{m1}\mathbf{S}^1_m\\
A_{12}\mathbf{S}^1_1+A_{22}\mathbf{S}^1_2+\dots +A_{m2}\mathbf{S}^1_m\\
\vdots\\
A_{1m}\mathbf{S}^1_1+A_{2m}\mathbf{S}^1_2+\dots +A_{mm}\mathbf{S}^1_m
\end{array} \right],\\
=&\frac{(1-\delta)(A_{1m}\mathbf{S}^1_1+A_{2m}\mathbf{S}^1_2+\dots +A_{mm}\mathbf{S}^1_m)}{det(I-\delta M)}\\
=&\frac{(1-\delta)D(\delta,\pi_1,\dots,\pi_{n},\mathbf{S}^1)}{det(I-\delta M)}
\end{aligned}.$$
Moreover,  followers have the same preference, since $S^{j_1}=S^{j_2}$, $j_1\neq j_2$.  For any $\pi_1$, we have $\pi_{-1}^{\overline{BR}}(\pi_1)\in \Pi_{-1}(\mathcal{B})$. 

Similar to the proof of Theorem  \ref{th::ZD-SSE-1}, player 1's best equalizer ZD strategy is $\pi_1^+$ or $\pi_1^-$. Then  $\max\limits_{\pi_1^{ZD}\in \Xi_1 (\delta)}U_1\left(\delta,\pi_1^{ZD}, \pi_{-1}^{\overline{BR}}(\pi_1^{ZD})\right)=L_2(\delta)$, and 
$$\min\limits_{\pi_1^{ZD}\in \Xi_1 (\delta)}U_1\left(\delta,\pi_1^{SSE}, \pi_{-1}^{\overline{BR}}(\pi_1^{SSE})\right)\!-\!U_1\left(\delta,\pi_1^{ZD}, \pi_{-1}^{\overline{BR}}(\pi_1^{ZD})\right) =
U_1^{SSE}-L_2(\delta).$$

\end{myproof}

Theorem  \ref{th::ZD-SSE-multi-compare} is an extension of Theorem \ref{th::ZD-SSE-1} in multi-player situations. It provides an upper bound of player $1$'s utility with equalizer ZD strategies, when facing many opponents with the same preference. Then the utility gap helps player $1$ make a tradeoff for an equalizer ZD strategy and the SSE strategy. On this basis, when setting the sum of opponents' utilities as $\Gamma_2^{+}(\delta)$ or $\Gamma_2^{-}(\delta)$, player 1 can receive the best utility of adopting equalizer ZD strategies. Also, computing the SSE strategy is difficult and time-consuming. Thus, player $1$ can choose the corresponding equalizer ZD strategy to avoid the cost of computing the SSE strategy, if it can tolerate the bounded loss.

\iffalse
 $U^1=\left[\begin{array}{cc} 
U^1_{11} & U^1_{12}\\
U^1_{21} & U^1_{22}\end{array}\right]=\left[\begin{array}{cc} 
4-(1-\theta)^2 & 1\\
3-(\theta-a)^2 & 2+(\theta-a)^2\end{array}\right]$, $U^2=\left[\begin{array}{cc} 
U^2_{11} & U^2_{12}\\
U^2_{21} & U^2_{22}\end{array}\right]=\left[\begin{array}{ll}
2.5 & 2\\
4 & 4\end{array}\right]$,
where $a$ is a parameter in the attacker's utility. 

\fi

\section{Examples}

We design experiments to verify that ZD strategy's maintenance of utility and action interaction,  compared with the baseline SSE strategy.

\subsection{Utility performance in UAVs}
We provide experiments to verify that the ZD strategy can help the leader maintain its utility, where the baseline is the SSE strategy. Let us consider a UAV problem among drones \cite{zhang2017strategic,zhang2020game}. 

\textbf{One-to-one scenario:} Here we consider a one-to-one scenario between player 1 and player 2. 
 Take $\theta\in[0,1]$ as a parameter in the game, $\delta =0.9$ and $U^i_{k,j}=p_{k,j}^i+q_{k,j}^i\theta^2$, where $p_{k,j}^i$ and  $q_{k,j}^i$ are uniformly generated in the
ranges $[0,4]$ and $[0,1]$, respectively, for $i\in N$. %We consider two cases, where $N=\{1,2\}$ for the two-player scenario, and $N=\{1,2,3\}$ for the three-player scenario. 
As for the strategy selection, we choose the equalizer ZD strategy according to Theorem \ref{th::ZD-SSE-1}, respectively, and solve the SSE through a mixed-integer non-linear program (MINLP) as shown in \cite{vorobeychik2012computing}.

\begin{figure}
\centering
\includegraphics[width=7cm]{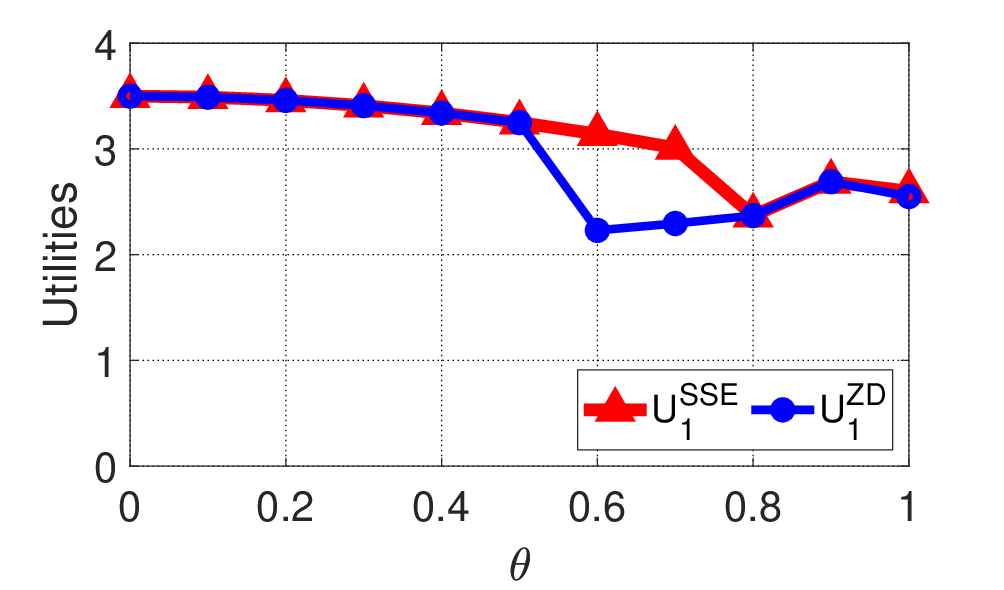}

\caption{Performance of the equalizer ZD strategy compared with the SSE strategy in the one-to-one situation. }
\label{fi::1}
\end{figure}

% Denote the optimal expected utility of players starting in state $s$ by $V_1(s)$ and $V_2(s)$. Consider the case where the current state is $s$, player $1$ takes strategy $\pi_1$, and player 2 takes action $b$. Players'  expected utilities are denoted by $\hat{r}_1(s,\pi_1,a)=\sum\limits_{a}\pi_1(a|s)\left(r_1(a,b)+\delta \sum\limits_{s'} P(s'|s,a,b) V_1(s) \right)$,$\hat{r}_2(s,\pi_1,a)=\sum\limits_{a}\pi_2(a|s)\left(r_2(a,b)+\delta \sum\limits_{s'} P(s'|s,a,b) V_2(s) \right)$. We solve the SSE through a mixed-integer non-linear program (MINLP) shown in \cite{vorobeychik2012computing}. 
\iffalse
\begin{equation}\label{eq::MILP}
\begin{aligned}
\max\limits_{\pi_1,\pi_2,a}  &U_1(\delta,\pi_1,\pi_2),\\
s.t. \ & \pi_1(a|s)\geqslant 0, \forall a, s, \sum\limits_1 \pi_1(a|s) = 1, \forall s,
\\
 & \pi_2(b|s)\in\{0,1\}, \forall b, s, \sum\limits_2 \pi_1(b|s) = 1, \forall s,
\\
&0\leqslant  V_2(s)-\hat{r}_2(\pi_1,b) \leqslant (1-\pi_2(b|s))M, \forall s%\!\forall i\!\in\mathbf{P}\!,k\!=\!1\!,\!\dots\!,\!K,
\\
&V_1(s)-\hat{r}_1(\pi_1,b) \leqslant (1-\pi_1(b|s))M, \forall s.%\!\forall i\!\in\mathbf{P}\!,k\!=\!1\!,\!\dots\!,\!K,
\end{aligned}
\end{equation}

\fi

In Fig \ref{fi::1}, we show the performance of the equalizer ZD strategy compared with the SSE strategy in the discounted repeated asymmetric game. Red dotted lines describe player 1's expected utilities adopting an SSE strategy, while blue solid lines describe player $1$'s expected utilities adopting the corresponding equalizer ZD strategy in Theorem \ref{th::ZD-SSE-1}. %We consider the two-player scenario and three-player scenario in Fig \ref{fig::1} and \ref{fig2}, respectively. 
As shown in Fig \ref{fi::1}, the expected utility of player 1 with an SSE strategy may be equal to the expected utility of player 1 with a ZD strategy, and is not lower than the expected utility of player 1 with a ZD strategy. Notice that the gap between utility with the SSE strategy and the ZD strategy may be zero, which is consistent with Theorem \ref{th::ZD-SSE-1}. Thus, player 1 can choose the corresponding equalizer ZD strategy to replace the SSE strategy to avoid computational resources.

\textbf{Three-player scenario:} In order to analyze the multiple-player situation, we consider a three-player scenario with player 1, player 2, and player 3. Similarly, take $\theta\in[0,1]$, $\delta =0.9$ and $U^i_{k,j}=p_{k,j}^i+q_{k,j}^i\theta^2$.  As for the strategy selection, we choose the equalizer ZD strategy according to Theorem \ref{th::ZD-SSE-multi-compare}.

\begin{figure}
\centering
\includegraphics[width=7cm]{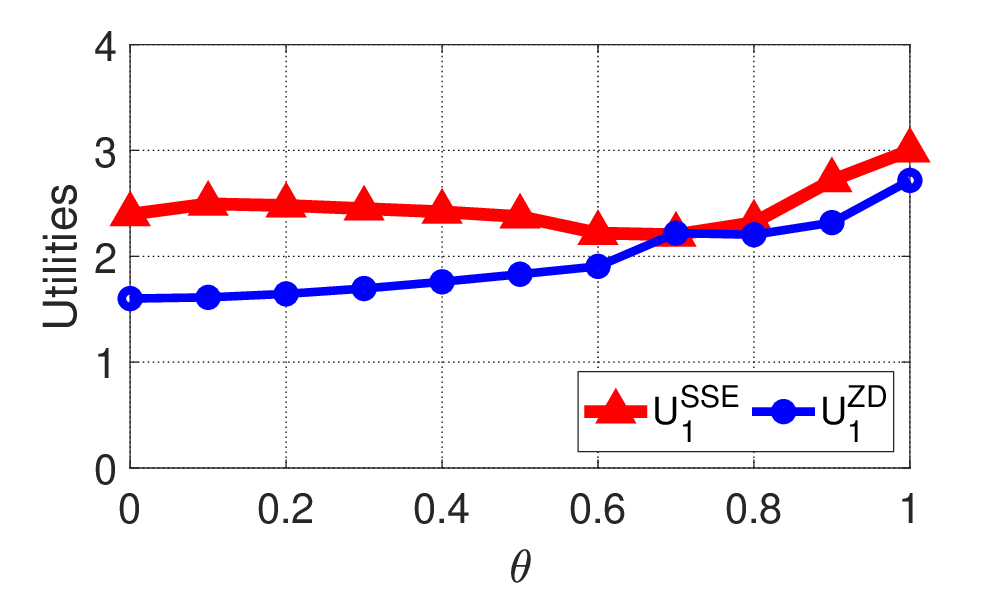}
\label{fig2}

\caption{Performance of the equalizer ZD strategy compared with the SSE strategy in the three-player situation. }
\label{fi::2}
\end{figure}

 As shown in Fig \ref{fi::2}, the expected utility of player 1 with an SSE strategy is always higher than that of player $1$  with a ZD strategy. Notice that the gap between utility with the SSE strategy and the ZD strategy is small, and is consistent with Theorem \ref{th::ZD-SSE-1}. Thus, if the gap is tolerable to player 1, player 1 can choose the corresponding equalizer ZD strategy to replace the SSE strategy in order to save computing resources.  

\subsection{Action Interaction in MTD}
In order to show the interaction process of the player, we specially simulated the selected actions in each stage. We consider MTD problems between a defender and attackers \cite{wang2019moving}.

\textbf{One-to-one scenario:} We consider a one-to-one scenario between player 1 (defender) and player 2 (attacker), while each player chooses action 1 (target 1) or action 2 (target 2).  We generate the utility matrix which satisfies
$\min\{U_{11}^1,U_{22}^1\}>\max\{U_{12}^1,U_{21}^1\}$ and $U_{11}^2<U_{12}^2$, and $U_{22}^2<U_{21}^2$. It means that the defender tends to protect the vulnerable target, and the attacker tends to implement invasions on the unprotected target.

\begin{figure}
\subfigure[Player $1$ selects the ZD strategy]{
\begin{minipage}[t]{1\linewidth}
\centering
\includegraphics[width=4in]{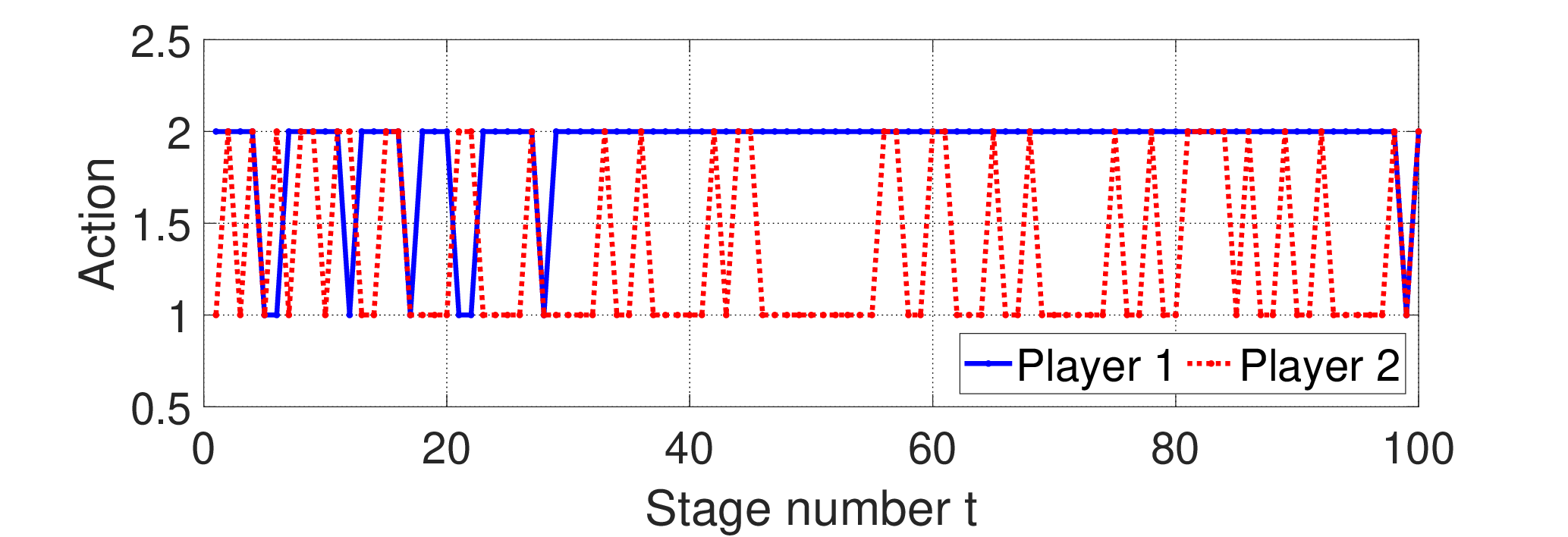}
%\caption{fig1}
\end{minipage}%
}\\
\subfigure[Player $1$ selects the SSE strategy]{
\begin{minipage}[t]{1\linewidth}
\centering
\includegraphics[width=4in]{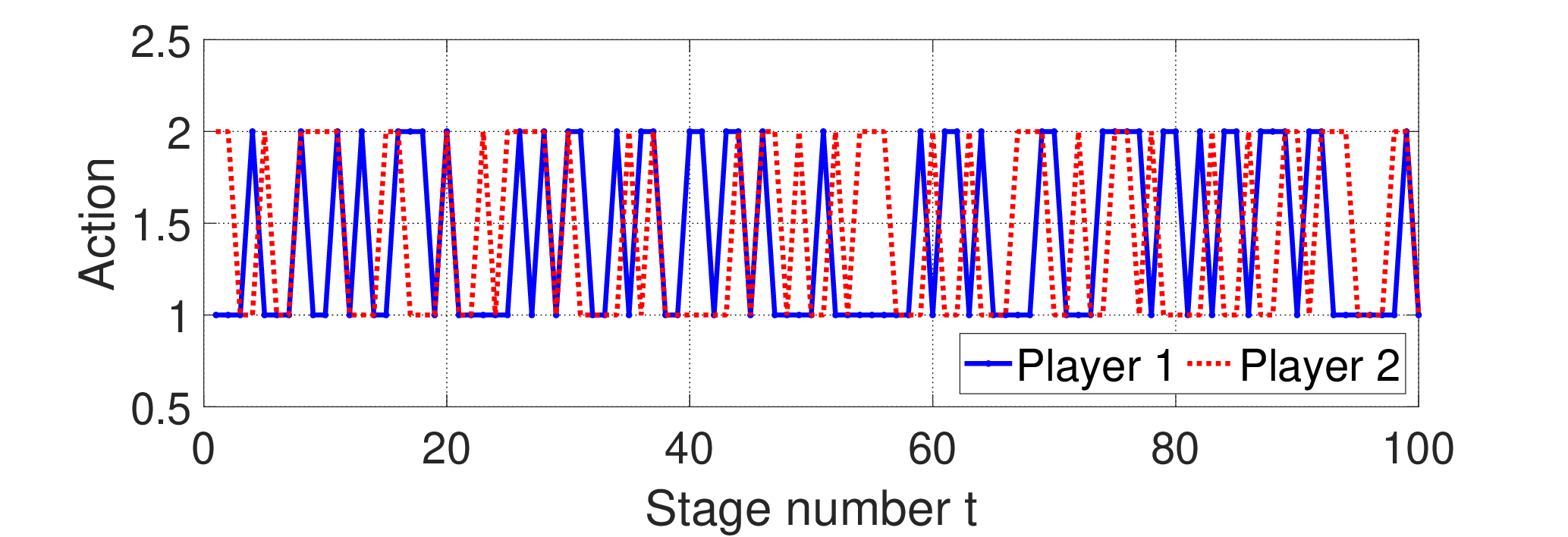}
%\caption{fig1}
\end{minipage}%
}%

\caption{Target interaction between player 1 (the defender) and player 2 (the attacker). Blue solid lines show the target that player 1 protects, while red dotted lines show the target that player 2 attacks in each stage. }
\label{fi::3}
\end{figure}

Moreover,  in Fig \ref{fi::3}(a),(b), we show the action interaction between the defender and the attacker. When the defender and the attacker choose the same target, the defender gains a high utility since it protects the right target under attack.  Notice that, although the SSE strategy has a better performance than the ZD strategy, the defender can still protect the right target  sometimes and get a good utility, which is consistent with Theorem \ref{th::ZD-SSE-1}.

\textbf{Three-player scenario:} Next, we consider a three-player scenario with player 1 (a defender), player 2 (an attacker), and player 3 (an attacker), while each player chooses action 1 (target 1) or action 2 (target 2).  We generates the utilities which stastify $U^1_{1,i}>U^1_{1,j}$, $U^1_{2,i}<U^1_{2,j}$, for $i<j$, and  $U^1_{1,i}<U^1_{1,j}$, $U^1_{2,i}>U^1_{2,j}$, for $i<j$ and $k\neq 1$. Thus, the defender hopes to select the targets that face many attacks, while the attacker has the opponent's preference as the defender.

\begin{figure}
\subfigure[Player $1$ selects the ZD strategy]{
\begin{minipage}[t]{1\linewidth}
\centering
\includegraphics[width=4in]{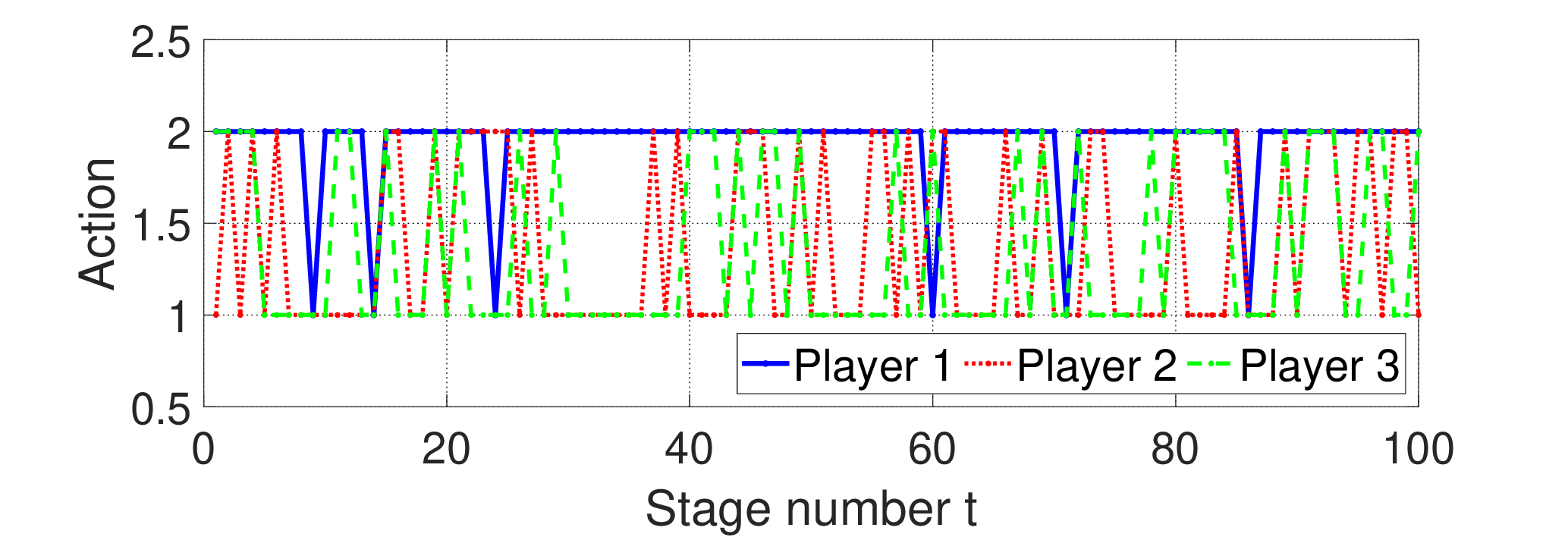}
%\caption{fig1}
\end{minipage}%
}%
\\
\subfigure[Player $1$ selects the SSE strategy]{
\begin{minipage}[t]{1\linewidth}
\centering
\includegraphics[width=4in]{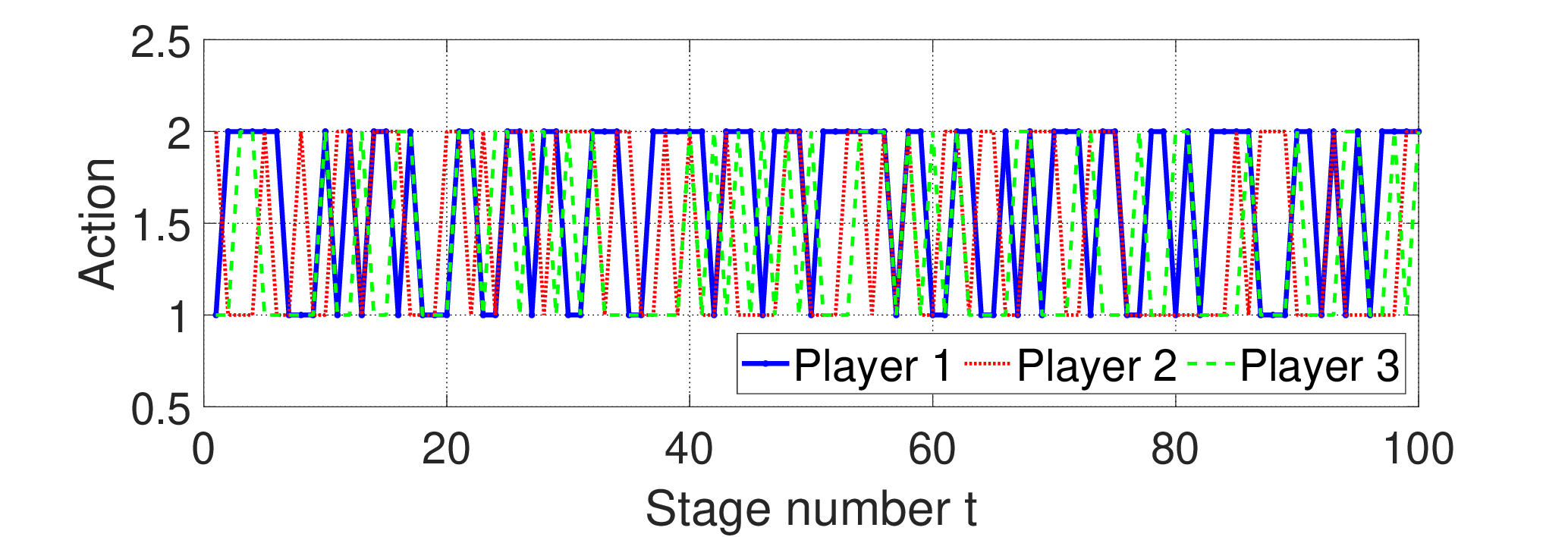}
%\caption{fig1}
\end{minipage}%
}%

\caption{Target interaction among player 1 (a defender), player 2 (an attacker), and player 3 (an attacker). Blue solid lines show the target that player 1 protects, while red dotted lines and green dotted lines show the target that player 2 and player 3 attack, respectively. }
\label{fi::4}
\end{figure}

In Fig \ref{fi::4}(a),(b), we show the action interaction among the three players.  Obviously, the defender, player $1$,  with the ZD strategy protects the target that the attacker invades frequently, which is not significantly worse than that with the SSE strategy. It is consistent with Theorem \ref{th::ZD-SSE-multi-compare}, that the ZD strategy has a good performance and brings bounded loss for player $1$ than the SSE strategy does.

\iffalse
For a player with a utility profile $\mathbf{f}=[f_{11},f_{12},f_{21},f_{22}]$ which describes the utility in the $4$ states, its expected utility can be expressed as 
$$U(\delta,\pi_d,\pi_1)=\frac{D(\delta,\pi_d,\pi_1,\mathbf{f})+(1-\delta)g(\delta,\pi_d,\pi_1,\mathbf{f})}{D(\delta,\pi_d,\pi_1,\mathbf{1})},$$
where $g(\delta,\pi_d,\pi_1,\mathbf{f})=\pi_1^0(1)\pi_1^0(1)D_1(\delta,\pi_d,\pi_1,\mathbf{f})+\pi_1^0(1)D_2(\delta,\pi_d,\pi_1,\mathbf{f})+\pi_1^0(1)D_3(\delta,\pi_d,\pi_1,\mathbf{f})$.
\fi

%\vspace{2mm}
\section{Conclusion} This paper studied equalizer ZD strategies in discounted repeated Stackelberg asymmetric games. In a typical case, the one-to-one situation, The existence condition of equalizer ZD strategies was verified, and an upper bound of the approach was revealed. Moreover, we extended our results into multi-player models and showed an upper bound with the equalizer ZD strategy. Finally, we gave a simulation of the interactions in UAVs and MTD problems to illustrate the effectiveness of our approach.

%\acknowledgements{\rm Thanks $\cdots$}
%% Please thank the anonymous people who make contributions to this article. If you don't want it, please delete it.

\bibliographystyle{IEEEtran}
\bibliography{wpref}

% Generated by IEEEtran.bst, version: 1.14 (2015/08/26)
\begin{thebibliography}{10}
\providecommand{\url}[1]{#1}
\csname url@samestyle\endcsname
\providecommand{\newblock}{\relax}
\providecommand{\bibinfo}[2]{#2}
\providecommand{\BIBentrySTDinterwordspacing}{\spaceskip=0pt\relax}
\providecommand{\BIBentryALTinterwordstretchfactor}{4}
\providecommand{\BIBentryALTinterwordspacing}{\spaceskip=\fontdimen2\font plus
\BIBentryALTinterwordstretchfactor\fontdimen3\font minus
  \fontdimen4\font\relax}
\providecommand{\BIBforeignlanguage}[2]{{%
\expandafter\ifx\csname l@#1\endcsname\relax
\typeout{** WARNING: IEEEtran.bst: No hyphenation pattern has been}%
\typeout{** loaded for the language `#1'. Using the pattern for}%
\typeout{** the default language instead.}%
\else
\language=\csname l@#1\endcsname
\fi
#2}}
\providecommand{\BIBdecl}{\relax}
\BIBdecl

\bibitem{liu2023optimal}
Y.~Liu and L.~Cheng, ``Optimal resource allocation and feasible hexagonal
  topology for cyber-physical systems,'' \emph{Journal of Systems Science and
  Complexity}, pp. 1--26, 2023.

\bibitem{chen2021distributed}
G.~Chen, Y.~Ming, Y.~Hong, and P.~Yi, ``Distributed algorithm for
  $\varepsilon$-generalized {Nash} equilibria with uncertain coupled
  constraints,'' \emph{Automatica}, vol. 123, p. 109313, 2021.

\bibitem{umsonst2020nash}
D.~Umsonst, S.~Sarita{\c{s}}, and H.~Sandberg, ``A {Nash} equilibrium-based
  moving target defense against stealthy sensor attacks,'' in \emph{Proceedings
  of the 59th IEEE Conference on Decision and Control (CDC)}.\hskip 1em plus
  0.5em minus 0.4em\relax IEEE, 2020, pp. 3772--3778.

\bibitem{xu2023algorithm}
G.~Xu, G.~Chen, and H.~Qi, ``Algorithm design and approximation analysis on
  distributed robust game,'' \emph{Journal of Systems Science and Complexity},
  vol.~36, no.~2, pp. 480--499, 2023.

\bibitem{miao2013stochastic}
F.~Miao, M.~Pajic, and G.~J. Pappas, ``Stochastic game approach for replay
  attack detection,'' in \emph{Proceedings of the 52nd IEEE Conference on
  Decision and Control (CDC)}.\hskip 1em plus 0.5em minus 0.4em\relax IEEE,
  2013, pp. 1854--1859.

\bibitem{zhang2018dynamically}
F.~Zhang, Z.~Zheng, and L.~Jiao, ``Dynamically optimized sensor deployment
  based on game theory,'' \emph{Journal of Systems Science and Complexity},
  vol.~31, pp. 276--286, 2018.

\bibitem{mishra2020model}
R.~K. Mishra, D.~Vasal, and S.~Vishwanath, ``Model-free reinforcement learning
  for stochastic {Stackelberg} security games,'' in \emph{Proceedings of the
  59th IEEE Conference on Decision and Control (CDC)}.\hskip 1em plus 0.5em
  minus 0.4em\relax IEEE, 2020, pp. 348--353.

\bibitem{feng2017signaling}
X.~Feng, Z.~Zheng, D.~Cansever, A.~Swami, and P.~Mohapatra, ``A signaling game
  model for moving target defense,'' in \emph{Proceedings of the 36th IEEE
  {C}onference on Computer Communications}.\hskip 1em plus 0.5em minus
  0.4em\relax IEEE, 2017, pp. 1--9.

\bibitem{li2020spatial}
H.~Li, W.~Shen, and Z.~Zheng, ``Spatial-temporal moving target defense: A
  markov {Stackelberg} game model,'' in \emph{Proceedings of the 19th
  International Conference on Autonomous Agents and MultiAgent Systems}, 2020,
  pp. 717--725.

\bibitem{tahir2019swarms}
A.~Tahir, J.~B{\"o}ling, M.-H. Haghbayan, H.~T. Toivonen, and J.~Plosila,
  ``Swarms of unmanned aerial vehicles—a survey,'' \emph{Journal of
  Industrial Information Integration}, vol.~16, p. 100106, 2019.

\bibitem{vorobeychik2012computing}
Y.~Vorobeychik and S.~Singh, ``Computing {Stackelberg} equilibria in discounted
  stochastic games,'' in \emph{Proceedings of the AAAI Conference on Artificial
  Intelligence}, vol.~26, no.~1, 2012, pp. 1478--1484.

\bibitem{korzhyk2011stackelberg}
D.~Korzhyk, Z.~Yin, C.~Kiekintveld, V.~Conitzer, and M.~Tambe, ``{Stackelberg}
  vs. {Nash} in security games: An extended investigation of
  interchangeability, equivalence, and uniqueness,'' \emph{Journal of
  Artificial Intelligence Research}, vol.~41, pp. 297--327, 2011.

\bibitem{cheng2023zero}
Z.~Cheng, G.~Chen, and Y.~Hong, ``Zero-determinant strategy in stochastic
  stackelberg asymmetric security game,'' \emph{Scientific Reports}, vol.~13,
  no.~1, p. 11308, 2023.

\bibitem{vasal2020stochastic}
D.~Vasal \emph{et~al.}, ``Stochastic {Stackelberg} games,'' \emph{arXiv
  preprint arXiv:2005.01997}, 2020.

\bibitem{9722864}
Z.~Cheng, G.~Chen, and Y.~Hong, ``Single-leader-multiple-followers
  {Stackelberg} security game with hypergame framework,'' \emph{IEEE
  Transactions on Information Forensics and Security}, vol.~17, pp. 954--969,
  2022.

\bibitem{lopez2022stationary}
V.~B. L{\'o}pez, E.~Della~Vecchia, A.~Jean-Marie, and F.~Ordonez, ``Stationary
  strong stackelberg equilibrium in discounted stochastic games,'' \emph{IEEE
  Transactions on Automatic Control}, 2022.

\bibitem{basu2023complexity}
A.~Basu, M.~Conforti, M.~Di~Summa, and H.~Jiang, ``Complexity of
  branch-and-bound and cutting planes in mixed-integer optimization,''
  \emph{Mathematical Programming}, vol. 198, no.~1, pp. 787--810, 2023.

\bibitem{basu2022complexity}
A.~Basu, ``Complexity of optimizing over the integers,'' \emph{Mathematical
  Programming}, pp. 1--42, 2022.

\bibitem{press2012iterated}
W.~H. Press and F.~J. Dyson, ``Iterated prisoner’s dilemma contains
  strategies that dominate any evolutionary opponent,'' \emph{Proceedings of
  the National Academy of Sciences}, vol. 109, no.~26, pp. 10\,409--10\,413,
  2012.

\bibitem{govaert2020zero}
A.~Govaert and M.~Cao, ``Zero-determinant strategies in repeated multiplayer
  social dilemmas with discounted payoffs,'' \emph{IEEE Transactions on
  Automatic Control}, vol.~66, no.~10, pp. 4575--4588, 2020.

\bibitem{tan2021payoff}
R.~Tan, Q.~Su, B.~Wu, and L.~Wang, ``Payoff control in repeated games,'' in
  \emph{Proceedings of the 33rd Chinese Control and Decision Conference
  (CCDC)}.\hskip 1em plus 0.5em minus 0.4em\relax IEEE, 2021, pp. 997--1005.

\bibitem{wang2016extortion}
Z.~Wang, Y.~Zhou, J.~W. Lien, J.~Zheng, and B.~Xu, ``Extortion can outperform
  generosity in the iterated prisoner's dilemma,'' \emph{Nature
  Communications}, vol.~7, no.~1, pp. 1--7, 2016.

\bibitem{hilbe2013evolution}
C.~Hilbe, M.~A. Nowak, and K.~Sigmund, ``Evolution of extortion in iterated
  prisoner’s dilemma games,'' \emph{Proceedings of the National Academy of
  Sciences}, vol. 110, no.~17, pp. 6913--6918, 2013.

\bibitem{hirai2013existence}
S.~Hirai and F.~Szidarovszky, ``Existence and uniqueness of equilibrium in
  asymmetric contests with endogenous prizes,'' \emph{International Game Theory
  Review}, vol.~15, no.~01, p. 1350005, 2013.

\bibitem{nockur2021different}
L.~Nockur, S.~Pfattheicher, and J.~Keller, ``Different punishment systems in a
  public goods game with asymmetric endowments,'' \emph{Journal of Experimental
  Social Psychology}, vol.~93, p. 104096, 2021.

\bibitem{reeves2017asymmetric}
T.~Reeves, H.~Ohtsuki, and S.~Fukui, ``Asymmetric public goods game cooperation
  through pest control,'' \emph{Journal of Theoretical Biology}, vol. 435, pp.
  238--247, 2017.

\bibitem{du2009asymmetric}
W.-B. Du, X.-B. Cao, M.-B. Hu, and W.-X. Wang, ``Asymmetric cost in snowdrift
  game on scale-free networks,'' \emph{Europhysics Letters}, vol.~87, no.~6, p.
  60004, 2009.

\bibitem{liang2015analysis}
H.~Liang, M.~Cao, and X.~Wang, ``Analysis and shifting of stochastically stable
  equilibria for evolutionary snowdrift games,'' \emph{Systems \& Control
  Letters}, vol.~85, pp. 16--22, 2015.

\bibitem{cheng2022misperception}
Z.~Cheng, G.~Chen, and Y.~Hong, ``Misperception influence on zero-determinant
  strategies in iterated prisoner’s dilemma,'' \emph{Scientific Reports},
  vol.~12, no.~1, pp. 1--9, 2022.

\bibitem{zhu2014promotion}
C.-J. Zhu, S.-W. Sun, L.~Wang, S.~Ding, J.~Wang, and C.-y. Xia, ``Promotion of
  cooperation due to diversity of players in the spatial public goods game with
  increasing neighborhood size,'' \emph{Physica A: Statistical Mechanics and
  its Applications}, vol. 406, pp. 145--154, 2014.

\bibitem{han2023complex}
J.-X. Han and R.-W. Wang, ``Complex interactions promote the frequency of
  cooperation in snowdrift game,'' \emph{Physica A: Statistical Mechanics and
  its Applications}, vol. 609, p. 128386, 2023.

\bibitem{zhang2023distributed}
H.~Zhang, G.~Chen, and Y.~Hong, ``Distributed algorithm for continuous-type
  {Bayesian} {Nash} equilibrium in subnetwork zero-sum games,'' \emph{IEEE
  Transactions on Control of Network Systems}, 2023.

\bibitem{chen2020learning}
G.~Chen, K.~Cao, and Y.~Hong, ``Learning implicit information in {Bayesian}
  games with knowledge transfer,'' \emph{Control Theory and Technology},
  vol.~18, pp. 315--323, 2020.

\bibitem{mutzari2021coalition}
D.~Mutzari, J.~Gan, and S.~Kraus, ``Coalition formation in multi-defender
  security games.'' in \emph{Proceedings of the AAAI Conference on Artificial
  Intelligence}, 2021, pp. 5603--5610.

\bibitem{sanjab2020game}
A.~Sanjab, W.~Saad, and T.~Ba{\c{s}}ar, ``A game of drones: Cyber-physical
  security of time-critical uav applications with cumulative prospect theory
  perceptions and valuations,'' \emph{IEEE Transactions on Communications},
  vol.~68, no.~11, pp. 6990--7006, 2020.

\bibitem{zhang2017strategic}
T.~Zhang and Q.~Zhu, ``Strategic defense against deceptive civilian gps
  spoofing of unmanned aerial vehicles,'' in \emph{Proceedings of the 8th
  International Conference on Decision and Game Theory for Security}.\hskip 1em
  plus 0.5em minus 0.4em\relax Cham: Springer, 2017, pp. 213--233.

\bibitem{zhang2020game}
T.~Zhang, L.~Huang, J.~Pawlick, and Q.~Zhu, ``Game-theoretic analysis of cyber
  deception: Evidence-based strategies and dynamic risk mitigation,''
  \emph{Modeling and Design of Secure Internet of Things}, pp. 27--58, 2020.

\bibitem{wang2019moving}
S.~Wang, H.~Shi, Q.~Hu, B.~Lin, and X.~Cheng, ``Moving target defense for
  internet of things based on the zero-determinant theory,'' \emph{IEEE
  Internet of Things Journal}, vol.~7, no.~1, pp. 661--668, 2019.

\end{thebibliography}

\end{document}